\documentclass[aps,prd,preprint,floatfix,11pt,showpacs,showkeys,tightenlines,nofootinbib,longbibliography,a4paper]{revtex4-1}
\usepackage[utf8]{inputenc}
\DeclareUnicodeCharacter{202F}{\,}
\usepackage{soul}
\usepackage[table]{xcolor}
\usepackage[compat=1.0.0]{tikz-feynman}
\usepackage{amsfonts}
\usepackage{amsmath}
\usepackage{array}
\usepackage[markup=nocolor]{changes}
\usepackage{graphicx}
\usepackage{hyperref}
\usepackage{multirow}
\usepackage{diagbox} 
\usepackage{commath}
\usepackage{booktabs}
\usepackage{url}
\usepackage{xspace}
\usepackage{setspace}
\usepackage{graphics}
\usepackage[parfill]{parskip}
\definecolor{RowGreen}{HTML}{28a745}
\definecolor{RowRed}{HTML}{dc3545}
\definecolor{RowOrange}{HTML}{fd7e14}
\definecolor{RowGray}{gray}{0.95}
\usepackage[most]{tcolorbox}
\newtcbox{\inlinebox}[2][]{enhanced,
 box align=base,
 nobeforeafter,
 colback=#2!20,
 colframe=#2!20,
 coltext=#2,
 boxrule=0.1pt,
  arc=1.8mm,
  auto outer arc,
  boxsep=2pt,
  left=2pt,
  right=2pt,
  bottom=0pt,
  top=0pt,
  #1}

\usepackage{amssymb}
\usepackage{pgfplots}
\hypersetup{pdfstartview=FitV,colorlinks=true,linkcolor=blue,citecolor=blue, filecolor=black,urlcolor=blue}
\definecolor{color1}{HTML}{cc9966}
\definecolor{color2}{HTML}{ff0000}
\definecolor{color3}{HTML}{dddd00}
\definecolor{color4}{HTML}{33ff00}
\definecolor{color5}{HTML}{ff00dd}
\definecolor{color6}{HTML}{0044dd}
\definecolor{color7}{HTML}{8822cc}
\definecolor{color8}{HTML}{00ffff}
\definecolor{color9}{HTML}{aa5511}
\definecolor{color10}{HTML}{9999cc}
\definecolor{color11}{HTML}{225522}
\definecolor{color12}{HTML}{ffa322}
\definecolor{sigstrcolor1}{HTML}{ee0000}
\definecolor{sigstrcolor2}{HTML}{00aaaa}
\definecolor{sigstrcolor3}{HTML}{550099}
\definecolor{sigstrcolor4}{HTML}{2255ff}
\definecolor{sigstrcolor5}{HTML}{22cc00}
\definecolor{sigstrcolor6}{HTML}{ff9900}
\definecolor{isocolor0}{HTML}{4fcf36}
\definecolor{isocolor12m}{HTML}{75cf36}
\definecolor{isocolor12}{HTML}{75cf36}
\definecolor{isocolor1}{HTML}{b4c400}
\definecolor{isocolor1m}{HTML}{b4c400}
\definecolor{isocolor32m}{HTML}{dddd00}
\definecolor{isocolor32}{HTML}{dddd00}
\definecolor{isocolor2}{HTML}{ddb800}
\definecolor{isocolor2m}{HTML}{ddb800}

\newcommand{\tr}{\mathrm{Tr}}
\newcommand{\be}{\begin{equation}}
\newcommand{\ee}{\end{equation}}
\newcommand{\bea}{\begin{eqnarray}}
\newcommand{\eea}{\end{eqnarray}}

\definecolor{lime}{HTML}{A6CE39}
\DeclareRobustCommand{\orcidicon}{\hspace{-1mm}
	\begin{tikzpicture}
	\draw[lime, fill=lime] (0,0) 
	circle [radius=0.16] 
	node[white] at (-0.007,-0.007) {{\fontfamily{qag}\selectfont \tiny \,ID}};
	\draw[white, fill=white] (-0.067,0.095) 
	circle [radius=0.005];
	\end{tikzpicture}
	\hspace{-3mm}
}

\foreach \x in {A, ..., Z}{\expandafter\xdef\csname orcid\x\endcsname{\noexpand\href{https://orcid.org/\csname orcidauthor\x\endcsname}
			{\noexpand\orcidicon}}
}
\usepackage{calc}
\newlength{\depthofsumsign}
\makeatletter
\newcommand*{\DivideLengths}[2]{%
  \strip@pt\dimexpr\number\numexpr\number\dimexpr#1\relax*65536/\number\dimexpr#2\relax\relax sp\relax
}
\makeatother

\begin{document}
\title{Light Scalars in the Extended Georgi-Machacek Model}
\author{Poulami Mondal\orcidB{}}
\altaffiliation{Corresponding author}
\email{\textcolor{black}{e-mail:} poulami.mondal@tifr.res.in}
\author{Subrata Samanta\orcidC{}}
\email{\textcolor{black}{E-mail:} samantaphys@gmail.com}
\affiliation{Department of Physics, Indian Institute of Technology Kanpur, Kanpur 208016, India}
\begin{abstract}
Searches for additional scalars belonging to the $SU(2)_L$ triplets at collider experiments make the Georgi-Machacek (GM) and extended GM (eGM) models particularly interesting due to their distinctive collider signatures while preserving $\rho=1$ at tree level. Specifically, the presence of a doubly-charged scalar in these models can radiatively enhance the partial decay width of a light neutral scalar decaying into two photons. We show that a Bayesian fit can simultaneously accommodate the ATLAS and CMS di-photon excesses together with the LEP $b\bar{b}$ excess around $95$~GeV in both the GM and eGM models, while being consistent with the measured properties of the $125$ GeV scalar at the LHC. In contrast, the CMS di-tau excess around $95$~GeV is compatible with both models only at the $\sim 2\sigma$ level, as its central value lies outside the region allowed by the 125~GeV Higgs signal strengths. The Bayesian fit also yields stringent upper bounds on the model parameters: (i) in the eGM model, the triplet vacuum expectation value (VEV) is constrained to
be approximately below 20 GeV, and the masses of the remaining scalars are bounded to be below 600 GeV with mass splittings not exceeding 250 GeV; (ii) in
the GM model, these constraints become even more stringent -- the triplet VEV is limited to be below 15 GeV, the custodial quintet and triplet scalar masses are constrained to be below 530 GeV and 320 GeV, respectively, and their mass splitting is restricted to be less than $210$ GeV. These results demonstrate that the combined impact of theoretical and experimental constraints leads to a relatively light scalar spectrum in the GM and eGM models, providing well-motivated frameworks for benchmark studies at both lepton and hadron colliders.
\end{abstract}
\date{\today}
\keywords{Beyond Standard Model, Higgs Physics, Multi-Higgs Models}
\maketitle
\section{Introduction}\label{sec:intro}
The Standard Model (SM) prediction of the production rate and decay branching fractions of the observed Higgs boson with a mass around $125$ GeV~\cite{ATLAS:2012yve,CMS:2012qbp} is well compatible with the Large Hadron Collider (LHC) measurements, leaving room for deviations of up to$\sim10\%$~\cite{ATLAS:2016neq,ATLAS:2020rej,ATLAS:2020qcv,ATLAS:2020fcp,ATLAS:2020fzp,CMS:2020xwi,ATLAS:2020bhl,CMS:2021ugl,CMS:2021kom,ATLAS:2021qou,ATLAS:2022yrq,CMS:2022ahq,CMS:2022kdi,CMS:2022uhn,CMS:2022dwd,ATLAS:2022vkf,ATLAS:2022ooq,ATLAS:2022tnm,CMS:2023tfj}.
Beyond the SM (BSM), the $SU(2)_L$ triplet scalars leave an imprint on the electroweak symmetry breaking (EWSB) and predict
large trilinear couplings of the new scalars to vector boson pairs, which mainly occur via triplet vacuum expectation value (VEV) and/or mixing with the SM Higgs boson. This can be tested through precision measurements of the 125 GeV Higgs boson couplings to vector boson pairs at lepton and hadron colliders. The search for BSM scalars is a primary focus of the current LHC programme, offering key insights into the mechanism behind EWSB.

In the ongoing search for an additional light Higgs-like particle at the LHC, mild excesses around 95~GeV have been reported in several channels. The CMS and ATLAS collaborations reported excesses in the di-photon final state~\cite{CMS:2015ocq,CMS:2018cyk,CMS:2024yhz,CMS-HIG-20-002,ATLAS:2024bjr}, and LEP reported an excess in the $b\bar{b}$ channel~\cite{LEPWorkingGroupforHiggsbosonsearches:2003ing,ALEPH:2006tnd}. CMS also reported an excess in the di-tau channel~\cite{CMS:2022goy}. The local significances of these excesses lie between about $2\sigma$ and $3\sigma$. The interpretation of the 95 GeV scalar excesses within the framework of extended scalar sector models has been the focus of numerous recent studies~\cite{Cacciapaglia:2016tlr,Cao:2016uwt,Fox:2017uwr,Liu:2018xsw,Domingo:2018uim,Biekotter:2019kde,Cline:2019okt,Kundu:2019nqo,Sachdeva:2019hvk,Cao:2019ofo,Abdelalim:2020xfk,Biekotter:2021qbc,Crivellin:2021ubm,Heinemeyer:2021msz,Biekotter:2022jyr,Biekotter:2022abc,Benbrik:2022azi,Iguro:2022dok,Iguro:2022fel,Kundu:2022bpy,Li:2022etb,Biekotter:2023jld,Azevedo:2023zkg,Escribano:2023hxj,Biekotter:2023oen,Belyaev:2023xnv,Ashanujjaman:2023etj,Aguilar-Saavedra:2023tql,Dutta:2023cig,Ellwanger:2023zjc,Cao:2023gkc,Borah:2023hqw,Ahriche:2023hho,Arcadi:2023smv,Ahriche:2023wkj,Chen:2023bqr,Dev:2023kzu,Cao:2024axg,Kalinowski:2024uxe,Ellwanger:2024txc,Ellwanger:2024vvs,YaserAyazi:2024hpj,Benbrik:2024ptw,Arhrib:2024wjj,Ge:2024rdr,Yang:2024fol,Lian:2024smg,Gao:2024qag,Khanna:2024bah,BrahimAit-Ouazghour:2024img,Gao:2024ljl,Kundu:2024sip,Mondal:2024obd,Banik:2024ugs,Hmissou:2025uep,Du:2025eop,Li:2025tkm,Xu:2025vmy,Benbrik:2025hol}.
The two-Higgs Doublet Model (2HDM), a minimal such extension, has been widely studied to explain these excesses~\cite{Cacciapaglia:2016tlr,Biekotter:2019kde,Heinemeyer:2021msz,Biekotter:2021qbc,Benbrik:2022azi,Biekotter:2022jyr,Belyaev:2023xnv,Biekotter:2023jld,Biekotter:2023oen,Benbrik:2024ptw,Banik:2024ugs,Khanna:2024bah,Xu:2025vmy,Benbrik:2025hol}. To account for both the LEP and LHC excesses within the 2HDM, charge-parity (CP) conservation in the scalar potential must be relaxed~\cite{Azevedo:2023zkg}. However, the parameter space accommodating both the CMS di-tau and di-photon excesses is in tension with flavour constraints, particularly from 
$b \rightarrow s\gamma$ transitions~\cite{Azevedo:2023zkg}. Models with $SU(2)_L$ triplet scalars feature a doubly-charged scalar that can enhance the loop-induced $\gamma\gamma$ decay rate of CP-even scalars,\footnote{Singly-charged scalars will also contribute, but the decay rate of a doubly-charged scalar is four times greater than that of a singly-charged scalar, assuming equal scalar couplings.} and explain the 95 GeV scalar excesses~\cite{Kundu:2022bpy,Iguro:2022fel,Ashanujjaman:2023etj,Ahriche:2023wkj,Chen:2023bqr,Dev:2023kzu,Kundu:2024sip,Du:2025eop,Li:2025tkm}. In addition to that, the scalar $SU(2)_L$ triplet extensions of the SM have various other interesting phenomenological aspects at the LHC~\citep{Georgi:1985nv,Chanowitz:1985ug,Gunion:1989ci,Gunion:1990dt,Aoki:2011pz,Kanemura:2012rs,Chiang:2012cn,Aoki:2012jj,Englert:2013zpa,Englert:2014uua,Hartling:2014zca,Garcia-Pepin:2014yfa,Hartling:2014aga,Chiang:2015amq,Chiang:2018xpl,Chiang:2018cgb,Banerjee:2019gmr,Ghosh:2019qie,Ashanujjaman:2021txz,deLima:2021llm,Kundu:2021pcg,Kanemura:2022ahw,Mondal:2022xdy,Chen:2022zsh,Ashanujjaman:2022ofg,Ghosh:2022bvz,Chen:2023ins,Chakraborti:2023mya,Das:2024xre,deLima:2024uwc,Crivellin:2024uhc,Chowdhury:2024mfu,Lu:2024ade,Ashanujjaman:2024lnr,Ashanujjaman:2025puu,Shang:2025cig,Lu:2025vif} and future colliders~\cite{10.21468/SciPostPhysProc.8.111,Bizon:2024juq,Stylianou:2023xit,Torndal:2023mmr,Celada:2024mcf,Aime:2022flm,MuonCollider:2022xlm,Forslund:2022xjq,Bhattacharyya:2025kvl,LinearColliderVision:2025hlt}.

Among the triplet scalar extensions of the SM, models that preserve $\rho=1$ at tree level are of particular interest.\footnote{Models with $\rho\neq 1$ at tree level are tightly constrained from electroweak precision data (see, for example, Ref.~\cite{Kanemura:2012rs}).} A minimal such extension of the SM requires one complex triplet with hypercharge $Y=1$
and one real triplet with $Y=0$, due to the custodial $SU(2)_V$ symmetry. The Georgi-Machacek (GM) model achieves this through a global $SU(2)_L\otimes SU(2)_R$ symmetry~\cite{Georgi:1985nv,Chanowitz:1985ug}. However, the global $SU(2)_L\otimes SU(2)_R$ is not necessary for constructing models with triplet scalars that preserve $\rho=1$ at tree level, as pointed out in Ref.~\cite{Kundu:2021pcg}. Additionally, Refs.~\cite {Gunion:1990dt,Englert:2013zpa,Hartling:2014aga,Blasi:2017xmc,Chiang:2018xpl,Keeshan:2018ypw,Chowdhury:2024mfu} showed that the global $SU(2)_L\otimes SU(2)_R$ symmetry needs to be broken to renormalise the GM model at the one-loop level. The extended GM (eGM) model, introduced in Ref.~\cite{Kundu:2021pcg}, maintains custodial $SU(2)_V$ symmetry at tree level without enforcing the global $SU(2)_L\otimes SU(2)_R$ symmetry in the scalar potential. Both the GM and eGM models yield $\rho=1$ at the tree level, while the eGM model possesses a more relaxed parameter space than the GM model and hence a richer collider phenomenology~\cite{Kundu:2021pcg,Chowdhury:2024mfu}.

Motivated by the recently reported excesses in the diphoton, $b\bar{b}$, and di-tau channels around 95~GeV, previous studies have investigated their interpretation within the GM model~\cite{Chen:2023bqr,Ahriche:2023wkj}. However, these analyses were performed using only tree level theoretical constraints.  Recently, in Ref.~\cite{Chowdhury:2024mfu}, we showed that next-to-leading-order (NLO) corrections to these constraints are important and significantly reduce the allowed parameter space of both the GM and eGM models. A comprehensive global analysis combining these improved theoretical constraints with the latest experimental constraints is therefore essential to identify the remaining viable parameter space of the GM and eGM models. In Ref.~\cite{Chowdhury:2024mfu}, we assumed all exotic scalars to be heavier than the observed 125~GeV Higgs boson. In this work, we instead identify the heavier CP-even custodial singlet with the SM-like Higgs boson $h$, with $m_h \approx 125$~GeV, and assume the lighter custodial singlet to have a mass around 95~GeV. We perform a Bayesian global fit of both the models, including NLO unitarity and bounded-from-below (BFB) conditions~\cite{Chowdhury:2024mfu},  the recent measurements of 95 GeV and 125 GeV Higgs signal strengths, and the latest direct search limits from heavy and light scalar searches at the LHC and LEP. Indirect limits from $B$-physics observables, as reported by HFLAV~\cite{HFLAV:2022esi}, are also included. These direct search and flavour constraints were not included
in Ref.~\cite{Chowdhury:2024mfu}, where the global fits were restricted to theoretical
constraints and the 125~GeV Higgs signal strengths.

In both the GM and eGM models, custodial symmetry is imposed only at tree level and is explicitly broken by the hypercharge gauge interactions. As a result, the one-loop correction to the $\rho$ parameter is ultraviolet divergent~\cite{Gunion:1990dt,Chowdhury:2026rus}. Its renormalisation, therefore, requires a counterterm associated with a custodial symmetry-breaking operator, which is absent from the tree level potential. Thus, the renormalisation of these models requires one additional independent input parameter beyond the three SM inputs $\alpha_{\rm em}$, $M_Z$, and $G_F$~\cite{Chowdhury:2026rus}. A comprehensive
global analysis of these models, incorporating an additional experimental input, is required to assess
the impact of the $T$ parameter on the GM and eGM model’s parameter space.
Although the $S$ and $U$ parameters are ultraviolet finite in both models~\cite{Hartling:2014aga,Drauksas:2023isd}, their experimental values are extracted from a global electroweak fit within the SM three-input scheme and are strongly correlated with $T$~\cite{deBlas:2016ojx,ParticleDataGroup:2024cfk}. A consistent treatment would therefore require a global electroweak fit within the four-input scheme, which we leave for future work.

From the global fit, we show that the LHC di-photon and LEP $b\bar{b}$ excesses around $95$ GeV are consistent with the measured properties of the $125$ GeV scalar at the LHC. Whereas the CMS di-tau excess around $95$~GeV is compatible with both models only at the $\sim 2\sigma$ level, as its central value lies outside the region allowed by the 125~GeV Higgs signal strengths. For the GM model, we obtain a stringent upper limit of approximately 15 GeV on the triplet VEV from the combined global fit, while the remaining scalar masses can reach up to about 530 GeV. Meanwhile, the eGM model allows for triplet VEVs up to 20 GeV and the remaining scalar masses up to around 600 GeV, thus providing well-motivated benchmark scenarios for searches for additional scalars at the LHC and future colliders.

The rest of the paper is organised as follows. A brief overview of these models is presented in Section~\ref{sec:model}. One-loop results of the $B$-physics observables are discussed in Section~\ref{sec:$B$-Physics}. Higgs signal strength observables are discussed in Section~\ref{sec:signal_strength}. We also briefly discuss the 95 GeV excesses reported by collider experiments and analyse the possibility of accommodating them within the GM and eGM models. We explain our global fit set-up and list all relevant constraints in Section~\ref{sec:fitting set-up and constraints}. The results from the global fit are presented in Section~\ref{sec:results}. Finally, in Section~\ref{sec:conclusion}, we summarise our findings and provide an outlook. In Appendix~\ref{app:A}, we list the 125~GeV Higgs signal strengths, the direct searches for light and heavy neutral BSM scalars, and the searches for light charged scalars in top-quark decays at the LHC used in our analysis. All the supplementary figures are placed in Appendix~\ref{app:C}.

\section{Models}\label{sec:model}
We extend the scalar sector of the Standard Model by introducing a real triplet $\xi$ with hypercharge $Y=0$ and a complex triplet $\chi$ with $Y=1$. The most general $SU(2)_{L}\otimes U(1)_{Y}$ invariant scalar potential, following the notation of Ref.~\cite{Chowdhury:2024mfu}, is given by,
\begin{align}
V&=-m_\phi^2\big(\phi^\dagger\phi\big)-m_\xi^2\big(\xi^\dagger\xi\big)-m_\chi^2\big(\chi^\dagger\chi\big)+\mu_1\big(\chi^\dagger t_a\chi\big)\xi_a+\mu_2\big(\phi^\dagger \tau_a\phi\big)\xi_a\nonumber\\
&\,\quad+\mu_3\Big[\big(\phi^T\epsilon\tau_a\phi\big)\tilde{\chi}_a+\text{h.c.}\Big]
+\lambda_\phi\big(\phi^\dagger\phi\big)^2+\lambda_\xi\big(\xi^\dagger\xi\big)^2+\lambda_\chi\big(\chi^\dagger\chi\big)^2\nonumber\\
&\,\quad+\tilde{\lambda}_\chi \big|\tilde{\chi}^\dagger\chi \big|^2+\lambda_{\phi\xi}\big(\phi^\dagger\phi\big)\big(\xi^\dagger\xi\big)
+\lambda_{\phi\chi}\big(\phi^\dagger\phi\big)\big(\chi^\dagger\chi\big)+\lambda_{\chi\xi}\big(\chi^\dagger\chi\big)\big(\xi^\dagger\xi\big)\nonumber\\
&\,\quad+\kappa_1\big|\xi^\dagger\chi \big|^2+\kappa_2\big(\phi^\dagger\tau_a\phi\big)\big(\chi^\dagger t_a\chi\big)+\kappa_3\Big[\big(\phi^T\epsilon\tau_a\phi\big)\big(\chi^\dagger t_a\xi\big)+\text{h.c.}\Big]\,.
\label{v16}
\end{align} 
Here, the scalar fields are defined as
$\phi=(\phi^+ \;\phi^0)^T$, $\xi=(\xi^+ \;\xi^0\;-\xi^{-})^T$, and $\chi=(\chi^{++} \;\chi^+ \;\chi^0)^T$,
where $\phi$ is the SM Higgs doublet, while $\xi$ and $\chi$ are the
real and complex triplet fields, respectively. The charge conjugate field associated with the field $\chi$ is defined as
$ \tilde{\chi}=(\chi^{0*} \;-\chi^{-} \;\chi^{--})^T$.
The matrices $\tau_a$ and $t_a$ denote the generators of $SU(2)$ in
the doublet and triplet representations, respectively. 
In this work, we consider a scalar potential with all parameters taken to be real, which ensures that CP is not explicitly violated.
After the EWSB, the neutral components of $\phi$, $\xi$, and $\chi$ acquire VEVs $v_\phi/\sqrt{2}$, $v_\xi$, and $v_\chi$, respectively. 
At tree level, the $\rho$ parameter is given by,
\begin{equation}
\nonumber
\rho=\frac{v_\phi^2+4\big(v_\xi^2+v_\chi^2\big)}{v_\phi^2+8 v_\chi^2}\,,\quad \text{with} \quad  v_\phi^2+4v_\xi^2+4v_\chi^2 = v^2 \approx (246 \, \text{GeV})^2\,. 
\end{equation}
Maintaining $\rho = 1$ at tree level requires $v_\chi = v_\xi$. 
Following~\cite{Kundu:2021pcg}, we ensure $\rho=1$ remains preserved by setting $v_{\chi}=v_{\xi}$ and imposing four constraints on the potential parameters, as outlined in Ref.~\cite{Chowdhury:2024mfu},
\begin{equation}
m_\chi^2=2m_\xi^2\,,\quad \mu_2=\sqrt{2}\mu_3\,,\quad \lambda_{\chi\xi}=2\lambda_\chi-4\lambda_\xi\,,\quad \kappa_2=4\lambda_{\phi\xi}-2\lambda_{\phi\chi}+\sqrt{2}\kappa_3\,.
\label{eq:constrints1}
\end{equation}
Note that these conditions are necessary to forbid any tree level mixing between the custodial singlet and non-custodial singlet states in the neutral sector, which ensures that the SM-like Higgs boson $h$ corresponds to a custodial singlet~\cite{Kundu:2021pcg,Chowdhury:2024mfu}. However, mixing among the singly charged scalars is still allowed at the tree level. Following~\citep{Kundu:2021pcg}, we refer to this framework as the extended GM (eGM) model. 

Apart from the would-be Goldstone bosons $G^\pm$ and $G^0$, the scalar spectrum consists of three CP-even neutral states, $F^0$, $H$, and $h$; one CP-odd neutral
state, $A$; two pairs of singly charged states, $H^\pm$ and $F^\pm$; and one pair of 
doubly charged states, $F^{\pm\pm}$. The mass-squared eigenvalues of the CP-even sector can be written as~\cite{Chowdhury:2024mfu},
\begin{align}
\nonumber
m_{h}^2=&\left(\mathcal{M}_n^2\right)_{11} c_\alpha^2 + \left(\mathcal{M}_n^2\right)_{22}  s_\alpha^2 -2\left(\mathcal{M}_n^2\right)_{12} s_\alpha c_\alpha\,,\\
\nonumber
m_{H}^2=&\left(\mathcal{M}_n^2\right)_{11} s_\alpha^2 + \left(\mathcal{M}_n^2\right)_{22}  c_\alpha^2 +2\left(\mathcal{M}_n^2\right)_{12} s_\alpha c_\alpha\,,
\\
m_{F^0}^2=&\frac{1}{2}v^2s_\beta^2\left(4\lambda_\xi-\lambda_\chi\right)-\frac{3}{2\sqrt{2}}v^2c_\beta^2\kappa_3+\frac{1}{\sqrt{2}}c_\beta^2s_\beta^{-1}\mu_2v+\frac{1}{\sqrt{2}}s_\beta\mu_1v\,,
\label{mass_cp_even}
\end{align}
where 
\begin{equation}\label{eq:mixing_angle_alpha}
\tan 2\alpha=\frac{2\left(\mathcal{M}_n^2\right)_{12}}{\left(\mathcal{M}_n^2\right)_{22}-\left(\mathcal{M}_n^2\right)_{11}}\,, \quad \text{with} \quad \alpha\in\left[-\frac{\pi}{2}\,,\frac{\pi}{2}\right]\,,
\end{equation}
and
\begin{align*}
\left(\mathcal{M}_n^2\right)_{11}&=2v^2c_\beta^2\lambda_\phi\,,\\
\left(\mathcal{M}_n^2\right)_{12}&=\frac{\sqrt{3}}{2}v^2s_{\beta}c_\beta\kappa_3+\sqrt{\frac{3}{2}}v^2s_{\beta}c_\beta\lambda_{\phi\xi}-\frac{\sqrt{3}}{2}c_\beta \mu_2 v\,,\\
\left(\mathcal{M}_n^2\right)_{22}&=v^2s_\beta^2\left(\lambda_\chi-\lambda_\xi\right)-\frac{1}{2\sqrt{2}}s_\beta\mu_1 v+\frac{1}{\sqrt{2}}c_\beta^2s_\beta^{-1}\mu_2v\,,
\end{align*}
with  $\tan\beta = 2 \sqrt{2} v_\chi/v_\phi$\,, where $c_\theta$ and $s_\theta$ stand for $\cos\theta$ and $\sin\theta$, respectively. We identify the
heavier CP-even custodial singlet state as the SM-like Higgs boson $h$ with mass $m_h\approx 125$ GeV, while the other custodial singlet state $H$ is assumed to have a mass below 125 GeV. The CP-odd neutral scalar $(A)$ and the doubly-charged scalar ($F^{++}$) mass-squared eigenvalues are as follows~\cite{Chowdhury:2024mfu},
\begin{align}
\nonumber
m_A^2&=\frac{1}{\sqrt{2}}s_\beta^{-1}\mu_2v-\frac{1}{2\sqrt{2}}v^2\kappa_3 \,,\\
m_{F^{++}}^2&=-\frac{1}{2\sqrt{2}}v^2c_\beta^2\left(\sqrt{2}\kappa_2+\kappa_3\right)+\frac{1}{2}v^2s_\beta^2\tilde{\lambda}_\chi+\frac{1}{\sqrt{2}}c_\beta^2s_\beta^{-1}\mu_2v+\frac{1}{\sqrt{2}}s_\beta\mu_1v\,.
\label{mass_cp_odd_doubly_charged}
\end{align}
The singly-charged mass-squared eigenvalues are given by~\cite{Chowdhury:2024mfu},
\begin{align}
\nonumber
m_{H^+}^2=&\left(\mathcal{M}_c^2\right)_{11} c_\delta^2 + \left(\mathcal{M}_c^2\right)_{22}  s_\delta^2 -2\left(\mathcal{M}_c^2\right)_{12} s_\delta c_\delta\,,\\
m_{F^+}^2=&\left(\mathcal{M}_c^2\right)_{11} s_\delta^2 + \left(\mathcal{M}_c^2\right)_{22}  c_\delta^2 +2\left(\mathcal{M}_c^2\right)_{12} s_\delta c_\delta\,,
\label{mass_singly_charged}
\end{align}
where
\begin{equation}
\tan 2\delta=\frac{2\left(\mathcal{M}_c^2\right)_{12}}{\left(\mathcal{M}_c^2\right)_{22}-\left(\mathcal{M}_c^2\right)_{11}}\,, \quad \text{with} \quad \delta \in\left[-\frac{\pi}{2},\frac{\pi}{2}\right]\,,
\label{mixing_angle_delta}
\end{equation}
and
\begin{align*}
\left(\mathcal{M}_c^2\right)_{11}&=-\frac{1}{8}v^2\left(\kappa_2+\sqrt{2}\kappa_3\right)+\frac{1}{\sqrt{2}}s_\beta^{-1}\mu_2v\,,\\ 
\left(\mathcal{M}_c^2\right)_{12}&=-\frac{1}{8}v^2c_\beta\left(\kappa_2-\sqrt{2}\kappa_3\right)\,,\\
\left(\mathcal{M}_c^2\right)_{22}&=-\frac{1}{8}v^2c_\beta^2\left(\kappa_2+5\sqrt{2}\kappa_3\right)+\frac{1}{4}v^2s_\beta^2\kappa_1+\frac{1}{\sqrt{2}}c_\beta^2s_\beta^{-1}\mu_2v+\frac{1}{\sqrt{2}}s_\beta\mu_1v\,.
\end{align*}
Imposing the global $SU(2)_L\otimes SU(2)_R$ symmetry on the potential in Eq.~(\ref{v16}) yields the following relations among the parameters:
\begin{align}\label{eq:GMconstraints}
m_\chi^2&=2m_\xi^2\,,&  \lambda_{\chi\xi}&=2\lambda_\chi-4\lambda_\xi\,,& \kappa_2&=4\lambda_{\phi\xi}-2\lambda_{\phi\chi}+\sqrt{2}\kappa_3\,,\nonumber\\
\mu_2&=\sqrt{2}\mu_3\,,& 
\kappa_2&=\sqrt{2}\kappa_3\,, & \kappa_1&=2\tilde{\lambda}_\chi=4\lambda_\xi-\lambda_{\chi\xi}\,.
\end{align}
Thus, Eq.~(\ref{v16}) reduces to the GM potential, which can be written in the notation of Ref.~\cite{Chiang:2015amq} as\footnote{Ref.~\cite{Chiang:2015amq} denotes the trilinear couplings by $\mu_1$ and $\mu_2$; we write them as $\tilde{\mu}_1$ and $\tilde{\mu}_2$.}
\begin{eqnarray}
\nonumber
V_{\text{tree}}(\Phi, \Delta) &=& 
\frac{1}{2}m^2_1 \tr[\Phi^\dagger \Phi]+\frac{1}{2}m^2_2 \tr[\Delta^\dagger\Delta]+\lambda_1(\tr[\Phi^\dagger \Phi])^2+ \lambda_2 (\tr[\Delta^\dagger \Delta])^2\nonumber \\
&&+ \lambda_3 \tr[(\Delta^\dagger \Delta)^2]+\lambda_4 \tr[\Phi^\dagger \Phi] \tr[\Delta^\dagger \Delta]+\lambda_5 \tr[\Phi^\dagger \tau^a \Phi \tau^b] \tr[\Delta^\dagger t^a \Delta t^b] \nonumber\\
&&+\tilde{\mu}_1 \tr[\Phi^\dagger \tau^a \Phi \tau^b](P^\dagger \Delta P)_{ab} + \tilde{\mu}_2 \tr[\Delta^\dagger t^a \Delta t^b](P^\dagger \Delta P)_{ab}\,,
\label{GM_pot}
\end{eqnarray}
where the SM doublet $\phi$ and the two triplets $\xi$ and $\chi$ are arranged into a bi-doublet $\Phi$ and a bi-triplet $\Delta$ of $SU(2)_L\otimes SU(2)_R$:
\begin{equation}
\Phi=
\begin{pmatrix}
\phi^{0*} & \phi^+\\
-\phi^- & \phi^0
\end{pmatrix},
\qquad
\Delta=
\begin{pmatrix}
\chi^{0*} & \xi^+ & \chi^{++}\\
-\chi^- & \xi^0 & \chi^+\\
\chi^{--} & -\xi^- & \chi^0
\end{pmatrix},
\label{GM_fields}
\end{equation}
where $a,b= 1,2,3$, and the matrix $P$ rotates $\Delta$ into the Cartesian basis,
\begin{equation}
    P = \begin{pmatrix} -\frac{1}{\sqrt2} & \frac{i}{\sqrt2} & 0  \\ 0& 0 & 1 \\ \frac{1}{\sqrt2} & \frac{i}{\sqrt2} & 0 \end{pmatrix}.
\end{equation}
Comparing Eqs.~(\ref{v16}) and (\ref{GM_pot}), we get
\begin{align}
    m_\phi^2&=-m_1^2\,, & m_\xi^2&=-\frac{1}{2}m_2^2\,, &  m_\chi^2&=-m_2^2\,, & \mu_1&=-6\tilde{\mu}_2\,,\nonumber\\
    \mu_2&=-\tilde{\mu}_1\,, & \mu_3&=-\frac{1}{\sqrt{2}}\tilde{\mu}_1\,, &  \lambda_\phi&=4\lambda_1\,, & \lambda_\xi&=\lambda_2+\lambda_3\,,\nonumber\\
    \lambda_\chi&=4\lambda_2+2\lambda_3\,, & \tilde{\lambda}_\chi&=2\lambda_3\,, &  \lambda_{\phi\xi}&=2\lambda_4\,, & \lambda_{\phi\chi}&=4\lambda_4\,,\nonumber\\
    \lambda_{\chi\xi}&=4\lambda_2\,, & \kappa_1&=4\lambda_3\,, &  \kappa_2&=2\lambda_5\,, & \kappa_3&=\sqrt{2}\lambda_5\,.\label{eq:eGMtoGM}
\end{align}
After EWSB, the neutral components of the bi-doublet and bi-triplet acquire VEVs,
\begin{equation}
\langle\Phi\rangle = \frac{v_\phi}{\sqrt2}\,\mathbb{I}_{2\times2}, \qquad
\langle\Delta\rangle = v_\Delta\,\mathbb{I}_{3\times3},
\end{equation}
with $v_\phi^2 + 8v_\Delta^2 = v^2 \approx (246~{\rm GeV})^2$. These vacuum configurations break the global $SU(2)_L\otimes SU(2)_R$ down to its diagonal subgroup, the custodial $SU(2)_V$. The diagonal form of $\langle\Delta\rangle$ implies equal VEVs for the real and complex triplets, $v_\xi = v_\chi \equiv v_\Delta$, which ensures $\rho = 1$ at tree level. The physical scalars are then classified into one fiveplet $(H_5^{\pm\pm}, H_5^\pm, H_5^0)$, one triplet $(H_3^\pm, H_3^0)$, and two singlets ($h$ and $H$) according to their transformation properties under the custodial $SU(2)_V$, and the states are identified with the eGM states as
\begin{equation*}
A \to H_3^0,\quad H^\pm \to H_3^\pm,\quad F^0 \to H_5^0,\quad F^\pm \to H_5^\pm,\quad F^{\pm\pm} \to H_5^{\pm\pm}, 
\end{equation*}
while the SM-like custodial singlet $h$ with $m_h \approx 125$~GeV remains unchanged. The 
relations in Eq.~(\ref{eq:GMconstraints}) ensure that, in the GM model, all physical scalars within a given custodial multiplet are mass-degenerate: $m_{H_3^0} = m_{H_3^\pm} \equiv m_3$ is the mass of the custodial triplet, $m_{H_5^0} = m_{H_5^\pm} = m_{H_5^{\pm\pm}} \equiv m_5$ is the mass of the custodial fiveplet, and $m_H$ denotes the mass of the other custodial singlet whose value is taken to be less than 125 GeV. The analytical expressions for these masses can be obtained from Eqs.~(\ref{mass_cp_even})--(\ref{mass_singly_charged}) by imposing the relations of Eq.~(\ref{eq:eGMtoGM}). For the custodial triplet and fiveplet, the masses are given by
\begin{equation}\label{eq:m5m3GM}
m_3^2=-\left(\frac{\tilde\mu_1}{4v_\Delta}+\frac{\lambda_5}{2}\right)v^2,\qquad
m_5^2=-\frac{\tilde\mu_1}{4v_\Delta}v_\phi^2-12\tilde\mu_2v_\Delta-\frac{3}{2}\lambda_5v_\phi^2+8\lambda_3v_\Delta^2 \,.
\end{equation}
For the two custodial singlets, $h$ and $H$, the masses are given by
\begin{align}
\nonumber
m_{h}^2=&\left(\mathcal{M}_n^2\right)_{11} c_\alpha^2 + \left(\mathcal{M}_n^2\right)_{22}  s_\alpha^2 -2\left(\mathcal{M}_n^2\right)_{12} s_\alpha c_\alpha\,,\\
m_H       ^2=&\left(\mathcal{M}_n^2\right)_{11} s_\alpha^2 + \left(\mathcal{M}_n^2\right)_{22}  c_\alpha^2 +2\left(\mathcal{M}_n^2\right)_{12} s_\alpha c_\alpha\,,\label{eq:mH1GM}
\end{align}
where 
\begin{align*}
\left(\mathcal{M}_n^2\right)_{11}&=8v^2c_\beta^2\lambda_1\,,\\
\left(\mathcal{M}_n^2\right)_{12}&=\sqrt{\frac{3}{2}}v^2s_{\beta}c_\beta(2\lambda_{4}+\lambda_5)+\frac{\sqrt{3}}{2}c_\beta \tilde\mu_1 v\,,\\
\left(\mathcal{M}_n^2\right)_{22}&=v^2s_\beta^2\left(3\lambda_2+\lambda_3\right)+\frac{3}{\sqrt{2}}s_\beta\tilde\mu_2 v-\frac{1}{\sqrt{2}}c_\beta^2s_\beta^{-1}\tilde{\mu}_1v\,,
\end{align*}
with $\tan\beta = 2\sqrt{2}\,v_\Delta/v_\phi$, and the mixing angle $\alpha$ given in Eq.~(\ref{eq:mixing_angle_alpha}).

In the GM and eGM models, fermion masses arise solely from the Higgs doublet $\phi$. Considering only the third generation of fermions, we have
\begin{equation}\label{eq:yukawa}
\mathcal{L}_Y = - Y_t \overline{Q}_L i\sigma_2 \phi^* t_R - Y_b \overline{Q}_L \phi b_R - Y_\tau \overline{L}_L \phi \tau_R + {\rm h.c.},
\end{equation}
with $Y_f = Y_f^{\rm SM}/\cos\beta$, since $v_\phi = v\cos\beta$. The physical scalars couple to fermions only through their doublet components. Hence, the fermionic couplings of the physical scalars  have the same structure as in the Type-I 2HDM~\cite{Branco:2011iw}.

The triplet field $\chi$ may also couple to lepton doublets through the lepton number-violating term $y_{ij}\,\overline{L^c_i}\, i\sigma_2\, \chi L_j$, which generates neutrino masses $m_\nu = 2\, y\, v_\chi$. For $v_\chi \sim \mathcal{O}(1~\textrm{GeV})$, $m_\nu \lesssim 0.1$~eV requires $y \lesssim 10^{-10}$. We therefore neglect the lepton number-violating Yukawa interactions between the lepton doublets and the triplet $\chi$. This approximation is valid for $v_\chi \gg 10^{-4}$~GeV, where the decays  $F^{++} \to \ell^+\ell^+$ (or $H_5^{++} \to \ell^+\ell^+$ in the GM model) are negligible compared with the di-boson channels~\cite{Chiang:2012cn}.

\section{$B$-Physics Observables}
\label{sec:$B$-Physics}
In the absence of tree level flavour-changing neutral current (FCNC) interactions, extended scalar sectors are constrained by loop-induced $B$-meson decays due to the presence of charged scalars~\cite{Hewett:1992is,Barger:1992dy,Buras:1993xp}. In both the GM and eGM models, the quark Yukawa interactions arise solely from the SM Higgs doublet (see Eq.~(\ref{eq:yukawa})). Among the charged scalars, only the singly-charged states ($H_3^\pm$ in the GM model; $H^\pm$ and $F^\pm$ in the eGM model) couple to quarks. As in the Type-I 2HDM, their couplings to both up- and down-type quarks are proportional to $\tan\beta$. In the absence of a $\tan\beta$ enhancement of the down-type couplings, the leading flavour constraints arise from loop-induced processes mediated by a top quark and a singly charged scalar. A detailed study of $b\to s\gamma$, $B_s^0$--$\bar B_s^0$ mixing, $B_s\to\mu^+\mu^-$, and the $Z$-pole observable $R_b$ in the GM model~\cite{Hartling:2014aga} has shown that $b\to s\gamma$ provides the most stringent bound on the triplet VEV. Therefore, in this paper, we focus on $B$-physics constraints, primarily arising from the quark-level transition $b\to s\gamma$. 
\subsection{ The Decays $B\to X_s\gamma$}
Constraints on exotic charged scalars from the $b\to s\gamma$ decay have been studied in~\cite{Hermann:2012fc,Misiak:2020vlo} and references therein for the 2HDM and in~\cite{Hartling:2014aga} for the GM model. Following~\cite{Misiak:2015xwa,Czakon:2015exa}, we consider two physical observables that impose the strongest constraints on the $b\to s\gamma$ amplitude: (i)  the CP- and isospin-averaged branching ratio $\mathcal{B}_{s\gamma}$ and (ii) the ratio,
\begin{equation}
R_\gamma\equiv \frac{\mathcal{B}_{(s+d)\gamma}}{\mathcal{B}_{c\ell\nu}}\,,
\end{equation}
where $\mathcal{B}_{(s+d)\gamma}$ and $\mathcal{B}_{c\ell\nu}$ are the inclusive branching ratios of the weak radiative and semileptonic decays. The current experimental average for $\mathcal{B}_{s\gamma}$ and $R_\gamma$ at $E_0= 1.6$ GeV read~\cite{Misiak:2017bgg,HFLAV:2022esi}
\begin{equation}
\mathcal{B}^{\text{exp}}_{s\gamma} = (3.49 \pm 0.19)\times  10^{-4}\,,\quad R^{\text{exp}}_\gamma = (3.22 \pm 0.15) \times 10^{-3}\,.
\label{eq:flavor_expt}
\end{equation}

\begin{figure}
  \centering
  \includegraphics[width=.8\linewidth]{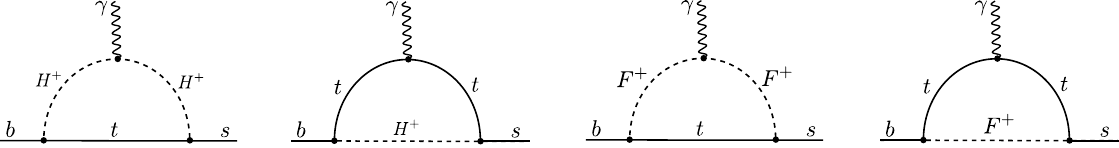}  
  \caption{One-particle irreducible (1PI) Feynman diagrams contributing to $\Delta C_7$ and $\Delta C_8$ at one-loop order in the eGM  model.}
  \label{fig:bsg_egm}
\end{figure}

Theoretical analyses of rare $B$-meson decays are conducted within an effective theory framework, where all heavier SM and relevant BSM particles are decoupled. In the presence of charged scalar particles, new physics contributions to the $b\to s\gamma$ transition are encoded in the Wilson coefficients $\Delta C_7$ and $\Delta C_8$,  which parameterise the new physics effects on the electromagnetic and chromomagnetic operators. Both the branching ratio $\mathcal{B}_{s\gamma}$ and the ratio $R_\gamma$ are sensitive to $\Delta C_7$ and $\Delta C_8$. They differ only in that $R_\gamma$ is normalised to the inclusive semileptonic rate, which reduces part of the theoretical uncertainty~\cite{Misiak:2017bgg}.
Their most general expressions are given up to Next-to-Next-to-Leading Order (NNLO) QCD corrections for the SM~\cite{Misiak:2006ab,Misiak:2006zs,Gunawardana:2019gep} and for the 2HDM~\cite{Hermann:2012fc}. Extending these results to the eGM model, which features two charged scalar particles coupling to fermions, is straightforward.
In this analysis, we consider the Leading Order (LO) contributions to the renormalised Wilson coefficients $\Delta C_7$ and $\Delta C_8$, arising from loop diagrams involving charged bosons. The dominant contributions come from diagrams where charged bosons couple to the top quark, as illustrated in Figure~\ref{fig:bsg_egm}. 
At the LO, the contributions of BSM charged scalars to the Wilson coefficients in the eGM model at the renormalisation scale $\mu_0$ are given by,
\begin{equation}
\Delta C_{7,8}(\mu_0)=\frac{1}{3}t^2_\beta c^2_\delta F_{7,8}^{(1)}(y_1)-t^2_\beta c^2_\delta F_{7,8}^{(2)}(y_1) +\frac{1}{3}t^2_\beta s^2_\delta F_{7,8}^{(1)}(y_2)-t^2_\beta s^2_\delta F_{7,8}^{(2)}(y_2)
\label{eq:deltac78}
\end{equation}
with
\begin{equation}
y_1\equiv\frac{\bar{m}^2_t(\mu_0)}{m_{H^\pm}^2},\quad y_2\equiv\frac{\bar{m}^2_t(\mu_0)}{m_{F^\pm}^2}\,.
\end{equation}
Here $\bar{m}_t$ represents the $\overline{MS}$ running mass, which is related to the pole mass $m_t$ by the relation~\cite{Ciuchini:1997xe},
\begin{equation}
\bar{m}_t(m_t)=m_t\left[1-\frac{4}{3}\frac{\alpha_s(m_t)}{\pi}\right],
\end{equation}
and the notations $t_\beta$, $s_\delta$, $c_\delta$ represent $\tan\beta$, $\sin\delta$, and $\cos\delta$, respectively. The NLO $t$-quark running mass at the scale $\mu_0$~\cite{Ciuchini:1997xe} is given by,
\begin{equation}
\bar{m}_t(\mu_0)=\bar{m}_t(m_t)\left[\frac{\alpha_s(\mu_0)}{\alpha_s(m_t)}\right]^{\frac{\gamma_0}{2\beta_0}}\left[1+\frac{\alpha_s(m_t)}{4\pi}\frac{\gamma_0}{2\beta_0}\left(\frac{\gamma_1}{\gamma_0}-\frac{\beta_1}{\beta_0}\right)\left(\frac{\alpha_s(\mu_0)}{\alpha_s(m_t)}-1\right)\right],
\end{equation}
\begin{equation}
\gamma_0=8,\quad \gamma_1=\frac{404}{3}-\frac{40}{9}n_f,\quad \beta_0=11-\frac{2}{3}n_f,\quad \beta_1=102-\frac{38}{3}n_f\,,
\end{equation}
where $n_f$ denotes the effective number of quark flavours and the values of $\alpha_s$ at a generic scale $\mu$ are computed using the NLO result from Ref.~\cite{Chetyrkin:1996vx,Ciuchini:1997xe},
\begin{equation}
\alpha_s(\mu)=\frac{\alpha_s(M_Z)}{r(\mu)}\left(1-\frac{\beta_1}{\beta_0}\frac{\alpha_s(M_Z)}{4\pi}\frac{\ln r(\mu)}{r(\mu)}\right),\quad r(\mu)=1-\beta_0\frac{\alpha_s(M_Z)}{2\pi}\ln\frac{M_Z}{\mu}\,.
\end{equation}
The forms of $F_{7,8}^{(1)}(y_i)$ and $F_{7,8}^{(2)}(y_i)$ for $i=1,2$ are given by~\cite{Hou:1987kf,Grinstein:1990tj,Ciuchini:1997xe},
\begin{align}\label{eq:wc78}
F_{7}^{(1)}(y_i)&=\frac{y_i(7-5y_i-8y_i^2)}{24(y_i-1)^3}+\frac{y_i^2(3y_i-2)}{4(y_i-1)^4}\ln y_i\,,\notag\\
F_{8}^{(1)}(y_i)&=\frac{y_i(2+5y_i-y_i^2)}{8(y_i-1)^3}-\frac{3y^2_i}{4(y_i-1)^4}\ln y_i\,,\notag\\
F_{7}^{(2)}(y_i)&=\frac{y_i(3-5y_i)}{12(y_i-1)^2}+\frac{y_i(3y_i-2)}{6(y_i-1)^3}\ln y_i\,,\notag\\
F_{8}^{(2)}(y_i)&=\frac{y_i(3-y_i)}{4(y_i-1)^2}-\frac{y_i}{2(y_i-1)^3}\ln y_i\,.
\end{align}
Incorporating improved non-perturbative contributions~\cite{Gunawardana:2019gep}, the updated SM predictions for $\mathcal{B}_{s\gamma}$ and $R_\gamma$, as estimated in~\cite{Misiak:2020vlo}, are given by, 
\begin{equation}
\mathcal{B}^{\text{SM}}_{s\gamma} = (3.40 \pm 0.17)\times  10^{-4}\,,\quad R^{\text{SM}}_\gamma = (3.35 \pm 0.16) \times 10^{-3}\,,
\end{equation}
for $E_0=1.6$ GeV. Thus, the effects of the additional contributions $\Delta C_{7,8}$ as given in Eq.~(\ref{eq:deltac78}) must be small; otherwise, they would increase the discrepancy between theory and experiment. In this scenario, $\mathcal{B}_{s\gamma}$ and $R_\gamma$ can be determined using the following simple linearised expressions~\cite{Misiak:2020vlo},
\begin{align}\label{eq:bsgammarange}
\mathcal{B}_{s\gamma} \times 10^4 &= (3.40 \pm 0.17)- 8.25\Delta C_7 (\mu_0) -2.10\Delta C_8 (\mu_0)\,,\notag\\
R_\gamma\times 10^3 &= (3.35 \pm 0.16) - 8.08\Delta C_7 (\mu_0) -2.06\Delta C_8 (\mu_0)\,,
\end{align}
where $\mu_0=160$ GeV. While the SM predictions in Eq.~(\ref{eq:bsgammarange}) include the full NNLO QCD corrections~\cite{Misiak:2020vlo}, the new-physics matching conditions $\Delta C_{7,8}(\mu_0)$ are evaluated at leading order in QCD. For Type-I 2HDM-like coupling structures, the NNLO QCD corrections to the charged-scalar matching conditions are moderate, and their main effect is to reduce the dependence on the matching scale $\mu_0$~\cite{Hermann:2012fc}. The observables $\mathcal{B}_{s\gamma}$ and $R_\gamma$ each impose an upper bound on the triplet VEV as a function of the charged scalar masses. The GM model contributions $\Delta C_{7,8}$ are derived from Eq.~(\ref{eq:deltac78}) by setting $\delta=0$ and considering the contribution of $H_3^\pm$ only. As shown in Section~\ref{sec:results}, the LHC direct searches provide stronger constraints on the triplet VEV than $\mathcal{B}_{s\gamma}$ in the low-mass region, so higher-order corrections to $\Delta C_{7,8}$ do not significantly affect the results of the global fit. 

\section{Higgs signal strength}\label{sec:signal_strength}
\subsection{SM-like Higgs, $h$}
The signal strength for the production of the SM-like Higgs boson $h$ (where $m_h = 125.09$ GeV) via production channel $i$ and its decay to final state $f$ is defined as,
\begin{equation}
\nonumber
\mu_i^f\equiv r_i\cdot \frac{r_f}{\displaystyle\sum_{\substack{f^\prime\in\{ ZZ, WW, Z\gamma,\\\gamma\gamma,  \mu\mu, bb,\tau\tau \}}}r_{f^\prime}\cdot \mathcal{B}_{\textrm{SM}}(h\to f^\prime)}\,,
\label{signal_strength_combined_h125}
\end{equation}
where $r_i$ and $r_f$ are the ratios of the production cross-sections $\sigma_i$'s (where $i\in\{gg$F$,\, bbh,\, $VBF$, $
$Wh, Zh,tth,th\}$) and the partial decay width $\Gamma^f$'s (where $f\in\{ ZZ,\, WW,\, \gamma\gamma,\, Z\gamma,\, \mu\mu,\, bb,\tau\tau \}$) to their corresponding SM values. $\mathcal{B}(h\to f)$'s are the decay branching fractions for final state $f$. \\[5pt]
In the $\kappa$-framework~\cite{ATLAS:2016neq}, the modifiers for SM Higgs coupling to vector bosons and fermions at tree level in both the GM and eGM models are given by,\footnote{Here we consider the SM-like Higgs $h$ as the heavy custodial singlet, while the mass of the other custodial singlet must be less than 125 GeV.}
\begin{equation}\label{eq:kappa_H125}
\kappa_{V}^h=s_\alpha c_\beta +\sqrt{\frac{8}{3}}c_\alpha s_\beta\,,\quad \textrm{and} \quad \kappa_{f}^h=\frac{s_\alpha}{c_\beta}\,, 
\end{equation}

The experimental input values of the $h$ signal strengths are taken from the latest CMS and ATLAS Run~2 data, summarised in Tables~\ref{tab:signalstrengthsC13} and~\ref{tab:signalstrengthsA13}, respectively. The combined ATLAS and CMS Run~1 measurements of the $h$ signal strengths are listed in Table~\ref{tab:signalstrengthsAC78}. All these tables are given in Appendix~\ref{app:A}.

\subsection{Light Higgs, $\Phi_{95}$}
Previously, LEP and, more recently, both Run 1 and Run 2 of the LHC have reported local excesses with a mass of around $95$ GeV~\cite{LEPWorkingGroupforHiggsbosonsearches:2003ing,ALEPH:2006tnd,CMS:2015ocq,CMS:2018cyk,CMS:2022goy,ATLAS:2024itc,CMS:2024yhz,CMS-HIG-20-002,ATLAS:2024bjr}. For a given final state $f$, the signal strength of the light Higgs boson with a mass of about $95$ GeV, $\Phi_{95}$ is defined as,
\begin{equation}
\nonumber
\mu_i^f\equiv r_i\cdot \frac{r_f}{\displaystyle\sum_{\substack{f^\prime\in\{ ZZ, WW, Z\gamma,\\\gamma\gamma,  \mu\mu, bb,\tau\tau \}}}r_{f^\prime}\cdot \mathcal{B}_{\textrm{SM}}(\Phi_{95}\to f^\prime)}\,.
\label{signal_strength_combined_h95}
\end{equation}
Here, $r_i$ and $r_f$ represent the ratios of the production cross-sections and partial decay widths, respectively, normalised to their SM expectations for a Higgs boson with a mass of approximately 95 GeV. $\mathcal{B}_{\textrm{SM}}(\Phi_{95}\to f^\prime)$ denotes the branching ratio of hypothetical SM Higgs boson $\Phi_{95}$ decaying into the final state $f^\prime$.

\subsubsection{The CMS and ATLAS di-photon excess}
Di-photon searches at the LHC are among the most promising avenues for detecting additional neutral scalars with masses below $125$ GeV. In recent years, the CMS collaboration has conducted searches for light neutral scalars decaying into two photons using LHC data at both 8 TeV~\cite{CMS:2015ocq} and 13 TeV~\cite{CMS:2018cyk}. By combining the Run 2 and Run 1 datasets, with integrated luminosities of $35.9\; \text{fb}^{-1}$ and $19.7\; \text{fb}^{-1}$, respectively, the CMS collaboration reported an excess consistent with a scalar resonance at a mass of $95.3$ GeV. This resonance, produced via gluon fusion and decaying into two photons, was reported with a local significance of $2.8\:\sigma$~\cite{CMS:2018cyk}. The updated CMS analysis with full Run 2 dataset with an integrated luminosity of $132\; \text{fb}^{-1}$ at 13 TeV, reported a  $2.9\:\sigma$ local excess at a mass of 95.4 GeV in the di-photon final state~\cite{CMS:2024yhz,CMS-HIG-20-002}. The corresponding signal strengths $\mu^{\gamma\gamma}_i$ for the production modes $i\in\{gg$F$,\,$VBF$\}$ are listed in Table~\ref{tab:95 signal strength}.\footnote{Notably, the updated CMS results report a significantly lower signal strength compared to previous analyses,  while the excess retains both its local significance and its preferred mass region. For further details, see~\cite{Azevedo:2023zkg}.}  On the other hand, the recent ATLAS analysis based on the full Run 2 dataset with an integrated luminosity of $140\; \text{fb}^{-1}$ at 13 TeV, reported a $1.7\:\sigma$ local excess in the di-photon final state at a mass near 95 GeV~\cite{ATLAS:2024bjr}. The reported local excess of events in the di-photon channel can be interpreted as a scalar resonance at a mass of 95.4~GeV, with a signal strength given in Table~\ref{tab:95 signal strength}.
\subsubsection{The CMS di-tau excess}
The CMS collaboration reported an excess over the expected background in light Higgs boson searches involving di-tau final states, consistent with a mass of $\sim95$ GeV under the assumption of gluon fusion production~\cite{CMS:2022goy}. The excess was most prominent at a mass of about 100 GeV, with global and local significances of  $2.7~\sigma$ and $3.1~\sigma$, respectively. It also remains consistent with a scalar resonance at 95.4 GeV, where the local and global significances are $2.6 ~\sigma$ and $2.3 ~\sigma$, respectively. The corresponding signal strength $\mu^{\tau\tau}_i$ is listed in Table~\ref{tab:95 signal strength}. To date, ATLAS has not reported an excess in the di-tau channel around $95$ GeV mass range.
\subsubsection{LEP excess}
The LEP reported a mild excess with a local significance of $2.3~ \sigma$ in the $e^+e^- \rightarrow Z(\Phi \rightarrow b\bar{b})$ searches~\cite{LEPWorkingGroupforHiggsbosonsearches:2003ing,ALEPH:2006tnd}, compatible with a scalar resonance at a mass near 95 GeV. The corresponding signal strength, $\mu^{bb}_i$, is given in Table~\ref{tab:95 signal strength}.
\begin{table}[ht]
\begin{center}
\setlength{\tabcolsep}{0pt}
\renewcommand{\arraystretch}{1.3}
\scalebox{0.85}{
\begin{tabular}{| c| ccc|ccc|}
\hline
 \textbf{Signal} & \;\multirow{2.4}{*}{\textbf{Value}}\; & \multicolumn{2}{c|}{\multirow{2.4}{*}{\;\;\textbf{Correlation matrix}\;}} & \;${\cal L}$\; & \multirow{2.4}{*}{\;\textbf{Detector}\;} & \multirow{2.4}{*}{\;\textbf{Source}\;}\\
\;  \textbf{strength} \;&  & \multicolumn{2}{c|}{} &\;\;\textbf{[fb$^{-1}$]}\;\; &  & \\[3pt]
\hline\hline
 $\mu_\text{ggF}^{\gamma \gamma}$ & $0.47 \pm 0.20$  &\;\;1 &\;$-0.52$ & \multirow{2.4}{*}{132.0} & \multirow{2.4}{*}{CMS} & \multirow{2.4}{*}{\,\cite{CMS:2024yhz,CMS-HIG-20-002} } \\[2pt]
 $\mu_\text{VBF}^{\gamma \gamma}$ & $0.06 \pm 0.70$  &\; \;$-0.52$ &\;1 & & & \\[4pt]
 $\mu_\text{pp}^{\gamma \gamma}$ & $0.18 \pm 0.10$  &  &  &140.0 & ATLAS & \cite{Biekotter:2023oen} \\
\hline
$\mu_\text{pp}^{\tau\tau}$ & $1.23 \pm 0.55$  & &&138.0 & CMS & \cite{CMS:2022goy} \\
\hline
$\mu_\text{ee}^{bb}$ & \;$0.117 \pm 0.057$\;  & &&2.461 & LEP & \cite{LEPWorkingGroupforHiggsbosonsearches:2003ing,Cao:2016uwt} \\
\hline
\end{tabular}}
\caption{Signal strength results of 95 GeV Higgs boson measured by ATLAS, CMS, and LEP. }
\label{tab:95 signal strength}
\end{center}
\end{table}
\subsubsection{GM/eGM interpretation of the 95 GeV excesses}
Here, we investigate the possibility of a neutral BSM scalar state with a mass of approximately 95 GeV within the CP-conserving GM and eGM models, assessing its ability to explain the excesses reported by CMS, ATLAS, and LEP. Both models feature three CP-even and one CP-odd scalar states, with two of the CP-even states being custodial singlets. We first examine whether a CP-even scalar state at 95 GeV can account for these excesses. Since the di-photon excess is primarily produced via gluon fusion, we exclude the possibility that a CP-even non-custodial singlet ($\Phi_{95}=H^0_5$ or $F^0$) is responsible for this excess, as it does not have any tree level couplings to fermions in these models. Therefore, in both models, we identify the lightest custodial singlet ($H$) as a new particle state with a mass of approximately 95 GeV, $\Phi_{95}=H$. The heavier custodial singlet ($h$) corresponds to the 125 GeV SM-like Higgs boson.

In the $\kappa$-framework~\cite{ATLAS:2016neq}, the modifiers for a hypothetical SM Higgs $H$ coupling to vector bosons and fermions at tree level in both the GM and eGM models are given by,
\begin{equation}\label{eq:kappa_h95}
\kappa^H_{V}=c_\alpha c_\beta -\sqrt{\frac{8}{3}}s_\alpha s_\beta\,,\quad \textrm{and} \quad \kappa^H_{f}=\frac{c_\alpha}{c_\beta}\,, 
\end{equation}
From Eqs.~(\ref{eq:kappa_H125}) and (\ref{eq:kappa_h95}), we obtain the following relations among the coupling modifiers: 
\begin{equation}
\kappa^h_{f}\kappa^h_{V}+\kappa^H_{f}\kappa^H_{V}=1\,.
\end{equation}


Next, we investigate whether the CP-odd scalar state ($\Phi_{95}=H^0_3$ or $A$) could be the source of the reported excess in both the CP-conserving GM and eGM models. The production of a scalar resonance via Higgsstrahlung at LEP requires a non-zero coupling of the scalar to the $Z$ boson. The interpretation of this excess as a CP-odd state in both models is disfavoured because of the vanishing coupling to vector bosons at tree level. Therefore, in this work, we exclude the CP-odd scalar state ($H^0_3$ or $A$) as a viable candidate for the 95 GeV mass resonance responsible for the reported excesses.

\section{Fit set-up and constraints}\label{sec:fitting set-up and constraints}
\subsection{Fit set-up}
To delineate the viable parameter space of the GM and eGM models, we perform a global analysis using the open-source code \texttt{HEPfit}~\cite{DeBlas:2019ehy}. This involves a Bayesian fit carried out with a Markov Chain Monte Carlo (MCMC) algorithm implemented through the \texttt{Bayesian Analysis Toolkit} (\texttt{BAT}) library~\cite{Caldwell:2008fw}. In our previous work~\cite{Chowdhury:2024mfu}, we extensively examined the viability of the GM and eGM models within the heavy Higgs scenario using the code \texttt{HEPfit}, where all BSM scalars are heavier than the observed 125 GeV Higgs boson. This work extends the parameter space to include the possibility where the BSM scalars are lighter than the 125 GeV Higgs boson.

In this study, we identify the heavier custodial singlet as the SM-like Higgs boson $h$ with a fixed mass of $m_h=125.09$ GeV, and vary the lighter custodial singlet ($H$) in the range $m_H \in [90,100]$ GeV. In our parametrisation, the quartic couplings are derived from the physical masses, the mixing angles, and the triplet VEV, so the experimental uncertainty $\delta m_h \simeq 0.24$~GeV~\cite{ParticleDataGroup:2016lqr} induces only a negligible shift in these couplings; for instance, $\lambda_1$ changes by $\mathcal{O}(10^{-5})$. The uncertainty on the Higgs mass, therefore, has no visible effect on the theoretical constraints or on the results of the global fit. All other Standard Model parameters are fixed to their best-fit values as reported in Ref.~\cite{deBlas:2016ojx}. In the case of the eGM model, we vary the other scalar masses within the range: $m_{F^0}, m_{F^{++}} \in[0.08,1]$ TeV. Due to stringent direct search limits on low-mass charged scalars from LEP~\cite{LEPHiggsWorkingGroupforHiggsbosonsearches:2001ogs}, we impose a lower bound of 80 GeV on the charged scalars in our analysis. We list the input parameters of the GM and the eGM models and their priors in Table~\ref{tab:GMeGMprior}. The prior ranges are chosen to be considerably larger than the regions allowed by
the fit (see Section~\ref{sec:results}), so that the upper limits of the allowed regions are set
by the theoretical and experimental constraints rather than by the priors. Following Refs.~\cite{Chowdhury:2017aav, Chiang:2018cgb, Chowdhury:2024mfu}, we
assign flat priors to all input parameters and carry out the global fits
separately with mass and mass-squared priors. The allowed regions shown in this
work correspond to the combination of the resulting 95.4\% probability regions. For further details on the implementation of the GM and eGM models in \texttt{HEPfit}, we refer the reader to Refs.~\cite{Chiang:2018cgb,Chowdhury:2024mfu}.
\begin{table}[!h]
\begin{center}
\setlength{\tabcolsep}{0pt}
\renewcommand{\arraystretch}{1.5}
\scalebox{1.}{
\begin{tabular}{| l ll|lll|}
\hline\hline
 \multicolumn{3}{|c|}{\bf GM} &  \multicolumn{3}{c|}{\bf eGM} \\
  {\;\small Parameters}\;\;&  &{\;\;\small Range}  & {\small \;Parameters\;} &  & {\;\;\small Range} \\
\hline
\;\;$v_\Delta$ &: & \;\;$ [0,60] $ GeV & \;\;$v_\chi$ & \;\;:& \;\;$ [0,60] $ GeV \\
\;\;$\alpha$  &: &  \;\;$[-\pi/2,\pi/2]\; $ & \;\;$\alpha,\delta$ &\;\;: & \;\;$[-\pi/2,\pi/2]$\; \\
\;\;$\lambda_3,\lambda_5$  &:& \;\;$[-4\pi,4\pi] $  & \;\;$\lambda_\chi$ &\;\;:& \;\;$[0,4\pi] $ \\
\;\;$m_H$  &:& \;\;$[90,100]$ GeV & \;\;$\tilde{\lambda}_\chi,\kappa_1,\kappa_3$ &\;\;: & \;\;$[-4\pi,4\pi] $ \\
\;\;$m_3,m_5$ &: & \;\;$[80,1000]$ GeV \;\; & \;\;$m_H$ &\;\;:& \;\;$[90,100]$ GeV\;\\
  & & & \;\;$m_{H^\pm},m_{F^\pm}$ &\;\;:& \;\;$[80,1000]$ GeV \;\;\\
\hline\hline
\end{tabular}}
\caption{Priors on the input parameters of the GM and eGM models.}
\label{tab:GMeGMprior}
\end{center}
\end{table}
\subsection{Theoretical constraints}
From a theoretical perspective, we incorporate the constraints given in Ref.~\cite{Chowdhury:2024mfu} in our fits: 
\begin{itemize}
\item The Higgs potential must satisfy stability bounds for all possible three field directions up to a scale of 1 TeV.\footnote{In our analysis, stability bounds refer to the scalar potential being bounded from below. The bound from metastability is not considered and will be explored in future work.}

\item Yukawa and quartic couplings must remain perturbative, bounded by $\sqrt{4\pi}$ and $4\pi$, respectively, up to a scale of 1 TeV. Their evolution is governed by two-loop renormalisation group equations. 

\item The $S$-matrix eigenvalues for $2\rightarrow 2$ scattering, including NLO corrections, must satisfy unitarity bounds at 1 TeV. We also require NLO corrections to be small relative to the LO eigenvalues.
\end{itemize}
\subsection{Experimental Constraints}
The experimental constraints included in our analysis are:
\begin{itemize}
\item The branching ratio $\mathcal{B}_{s\gamma}$ and the ratio $R_\gamma$.
\item $h$ signal strengths from
the latest ATLAS and CMS Run 2 data at a centre-of-mass energy of 13 TeV (Tables~\ref{tab:signalstrengthsC13} and~\ref{tab:signalstrengthsA13}) and ATLAS and CMS combination for Run 1 data (Table~\ref{tab:signalstrengthsAC78}).
\item $\Phi_{95}$ signal strengths given in Table~\ref{tab:95 signal strength}.
\item Total decay widths of $h$ and $t$-quark given in Table~\ref{tab:decaywidths}.
\item Direct search limits of the heavy and light BSM scalars at the LHC (Tables~\ref{tab:direct_search},~\ref{tab:ATLAS_light}--\ref{tab:ATLAS_charged}).\footnote{In the heavy scalar scenario, BSM scalars are heavier than the observed 125 GeV scalar, while the light scalar scenario corresponds to the BSM scalars with masses below 125 GeV.}
\end{itemize}
\begin{table}[ht]
\begin{center}
\setlength{\tabcolsep}{0pt}
\renewcommand{\arraystretch}{1.3}
\scalebox{0.85}{
\begin{tabular}{| c| c|ccc|}
\hline
\;\;\textbf{Observable}\;\; & \textbf{Value}  & \;\;$\cal L\; \textbf{[fb$^{-1}$]} $ \;\;& \;\;\textbf{Detector} \;\;&\;\; \textbf{Source}\;\;\\
\hline\hline
 \multirow{2}{*}{$\Gamma_h$} & $3.20 \pm 2.05$ MeV  & 138.0& CMS &\cite{CMS:2022dwd}  \\[2pt]
     &$4.50 \pm 2.75$ MeV  & 139.0& ATLAS &\cite{ATLAS:2023dnm}  \\[2pt]
     \hline
 $\Gamma_t$ & \;\;$1.360 \pm 0.127$  GeV\;\; & 19.7 & CMS &\cite{CMS:2014mxl}  \\[2pt]
\hline
\end{tabular}}
\caption{Total decay widths of SM Higgs boson and top quark.}
\label{tab:decaywidths}
\end{center}
\end{table}
\begin{table}[!ht]
\begin{center}
\setlength{\tabcolsep}{0pt}
\renewcommand{\arraystretch}{1.3}
\scalebox{1.}{
\begin{tabular}{| c| ccc|}
\hline
\multirow{2.4}{*}{\textbf{Channel}} & \multirow{2.4}{*}{\;\;\textbf{Experiment}\;\;} &\; \;\textbf{Mass range}\;\;\; & \;\;${\cal L}$\; \\
&  &\textbf{[TeV]}    &\;\;\textbf{[fb$^{-1}$]}\;\; \\
\hline\hline
\multirow{2}{*}{$pp\to H^{\pm},F^{\pm} \to \tau^\pm \nu $} & ATLAS~\cite{ATLAS:2014otc} & [0.18;1] &  {19.5}\\
&   CMS~\cite{CMS:2015lsf} &  [0.18;0.6] &   {19.7}\\
\hline
\multirow{2}{*}{$pp\to H^{\pm},F^{\pm} \to \tau^\pm \nu $} & ATLAS~\cite{ATLAS:2018gfm} & [0.2;2] &  {36.1}\\
&   CMS~\cite{CMS:2019bfg} &  [0.08;3] &   {35.9}\\
\hline
\multirow{2}{*}{$pp\to H^\pm ,F^\pm \to t \bar b$} & CMS~\cite{CMS:2020imj} & [0.2;3] &  {35.9}\\
&   ATLAS~\cite{ATLAS:2021upq} &  [0.2;2] &   {139.0}\\
\hline
\multirow{2}{*}{$pp\to H^+,F^+ \to t\bar b$} & CMS~\cite{CMS:2015lsf} & [0.18;0.6] &  {19.7}\\
&   ATLAS~\cite{ATLAS:2015nkq} &  [0.2;0.6] &   {20.3}\\
\hline
\multirow{2}{*}{$pp\to\tilde{\ell}^+\tilde{\ell}^-\to \ell^+\ell^-+2\tilde{\chi}^0$} & CMS~\cite{CMS:2024gyw} & [0.1;0.8] &  {137.0}\\
&  ATLAS~\cite{ATLAS:2022hbt} &  [0.09;0.35] &   {139.0}\\
\hline
\multirow{2}{*}{$pp\to\tilde{\tau}^+_L\tilde{\tau}^-_L\to \tau^+\tau^-+2\tilde{\chi}^0$} & CMS~\cite{CMS:2022syk} & [0.09;0.5] &  {138.0}\\
&  ATLAS~\cite{ATLAS:2019gti} &  [0.08;0.45] &   {139.0}\\
\hline
\;\;$e^+e^-\to\tilde{l}^+\tilde{l}^-\to l^+l^-+2\tilde{\chi}^0$\;\; & LEP~\cite{LEP-SUSY} & [0.045;0.102] &  {0.68}\\
\hline
\;\;$WZ\to H^+,F^+ \to WZ\:[\to(qq)(\ell\ell)]$\;\; & ATLAS~\cite{ATLAS:2015edr} & [0.2;1] &  {20.3}\\
\hline
\multirow{2}{*}{\;$WZ\to H^+,F^+ \to WZ\:[\to(\ell\nu)(\ell\ell)]$\;} & CMS~\cite{CMS:2021wlt} & [0.2;3] &  {137.0}\\
&   ATLAS~\cite{ATLAS:2024txt} &  [0.2;3] &   {140.0}\\
\hline
$VV\to F^{++} \to WW\:[\to(\ell\nu)(\ell\nu)]$ & CMS~\cite{CMS:2014mra} & [0.2;0.8] &  {19.4}\\
\hline
\multirow{2}{*}{$VV\to F^{++} \to WW\:[\to(\ell\nu)(\ell\nu)]$} & CMS~\cite{CMS:2021wlt} & [0.2;3] &  {137.0}\\
& ATLAS~\cite{ATLAS:2024txt} & [0.2;3] &  {140.0}\\
\hline
$pp\to F^{\pm\pm}F^{\pm\pm} \to W^{\pm}W^{\pm}W^{\pm}W^{\pm}$ & ATLAS~\cite{ATLAS:2021jol} & [0.2;0.6] &  {139.0}\\
\hline
\end{tabular}}
\caption{List of direct searches in LHC and LEP for singly- and doubly-charged scalars in the eGM model. Parentheses denote the final state decays of SM particles produced from the decays of BSM particles. In slepton searches, $l$ denotes all three charged lepton flavours, while $\ell$ refers specifically to electrons and muons.}
\label{tab:direct_search}
\end{center}
\end{table}
In this work, we analyse the eGM model, incorporating constraints from direct searches for heavy and light BSM scalars at the LHC. Furthermore, we include the latest exclusion limits from direct searches for both heavy and light BSM scalars for the GM model, providing an updated comparison with previous fits~\cite{Chiang:2018cgb}. These direct search constraints are implemented as follows:
\begin{enumerate}
\item For a given process, $X\to \Phi \to Y$,  we compute the theoretical production cross-section of a scalar particle $\Phi\: (=A,F^0,H^\pm,F^\pm,F^{\pm\pm})$, and multiply it by its branching fraction to a specific decay mode, i.e., $(\sigma\cdot \mathcal{B})_{\tiny\textrm{th}}=\sigma_{\tiny\textrm{th}}(X\to\Phi)\times\mathcal{B}_{\tiny\textrm{th}}(\Phi\to Y)$.\\
\item Following~\cite{Cacchio:2016qyh,Chowdhury:2017aav,Coutinho:2024zyp}, we define the ratio, 
\begin{equation}
R_{\tiny\textrm{direct}}=\frac{(\sigma\cdot \mathcal{B})_{\tiny\textrm{th}}}{(\sigma\cdot \mathcal{B})_{\tiny\textrm{obs}}}\,,
\end{equation}
where $(\sigma\cdot\mathcal{B})_{\rm th}$ is the theoretical prediction for the production cross section times branching ratio, and $(\sigma\cdot\mathcal{B})_{\rm obs}$ is the observed 95\% CL upper limit reported by the experimental collaboration. The experimental limits are provided at discrete mass points, which we linearly interpolate to obtain a continuous function of the scalar mass. Since $R_{\rm direct} \geq 0$, we assign to it a normal distribution centred at zero and restricted to positive values. This yields a one-sided constraint, as appropriate for an upper limit. The standard deviation is fixed by requiring that $R_{\rm direct} = 1$ is excluded at 95\% probability, which gives $\sigma  \simeq 0.51$. Additionally, since the experimental collaborations do not provide the correlations between different searches, each search channel is treated as an independent constraint in the global fit.
\end{enumerate}

Among the neutral CP-even scalars, the SM-like state $h$ is fixed at $m_h = 125.09$~GeV, while the lighter custodial singlet $H$ (with mass $m_H$ in the eGM and $m_1$ in the GM model) is varied in $[90,100]$~GeV. The remaining CP-even state $F^0$ ($H_5^0$) is fermiophobic, and the CP-odd state $A$ ($H_3^0$) is gauge-phobic at tree level. The theoretical predictions of $\sigma\cdot\mathcal{B}$ for the neutral and singly-charged scalars are obtained from the production cross sections computed in the (Aligned-) 2HDM~\cite{Cacchio:2016qyh,Chowdhury:2017aav,Coutinho:2024zyp}
for unit coupling modifiers, which are then rescaled to the GM/eGM model using the corresponding coupling modifiers $\kappa_i^2\;(i=f,W,Z)$. For example, the gluon-fusion and
$b\bar b/t\bar t$-associated production cross sections scale as $\kappa_f^2$,
while the VBF and $VH$ cross sections scale as $\kappa_W^2$ or $\kappa_Z^2$. The branching ratios are computed within the
GM/eGM model, including decays into lighter BSM scalars. Similarly, the cross sections of the fermion-induced production modes, $pp \to tbH^\pm$ and $t \to H^+ b$, are obtained from the 2HDM or Aligned-2HDM by rescaling with the corresponding fermionic coupling modifiers, $\kappa_f^{2}$. The $W^\pm Z$ fusion production of the charged scalar, which is absent in the (Aligned-) 2HDM, is taken from the GM implementation in \texttt{HEPfit}~\cite{Chiang:2018cgb} and weighted by $s_\delta^2$ for $H^\pm$ and $c_\delta^2$ for $F^\pm$ in the eGM model. For the doubly-charged scalar, we use the theoretical predictions of $\sigma\cdot\mathcal{B}$ implemented in \texttt{HEPfit} for the GM model~\cite{Chiang:2018cgb}. For the eGM model, the production cross sections apply equally, whereas the branching ratios are computed within the eGM model, including its cascade decays into lighter BSM scalars. This is because the $F^{\pm\pm}$ is a pure component of the complex triplet $\chi$ and has identical gauge boson couplings. Theoretical uncertainties on $\sigma\cdot\mathcal{B}$ are not included in the fits.

For the singly-charged scalars, we also consider LEP and LHC searches for sleptons. A slepton decaying into a charged lepton and a neutralino produces the same collider signature as a charged scalar decaying into a charged lepton and a neutrino, provided the neutralino is nearly massless. We therefore include the slepton searches listed in Table~\ref{tab:direct_search} and reinterpret their limits, taken in the massless-neutralino limit, as constraints on $H^\pm$ and $F^\pm$ (on $H_3^\pm$ and $H_5^\pm$ in the GM model). Only searches for left-handed sleptons are used, since their pair production proceeds through the same $\gamma/Z$ couplings as that of the charged scalars. The Drell--Yan pair-production cross sections of left-handed sleptons are taken from Ref.~\cite{Freitas:2003yp} for the LEP and from Refs.~\cite{LHC-SUSY-WG,Fiaschi:2018xdm,Fuks:2013lya} at NLO+NLL for the LHC.

The presence of light scalars can modify the (invisible) total decay widths of several well-measured SM particles, such as the Higgs boson and the top quark. For instance, if the masses of BSM neutral scalars, $\Phi=H$ or $A$ in the eGM model ($\Phi=H$ or $H_3^0$ in the GM model), are lighter than the SM-like Higgs $h$, the cascade decays $h\to \Phi\Phi^{*}\to \Phi b\bar{b}$ can be important in the region with low triplet VEV ($v_\chi$ in the eGM and $v_\Delta$ in the GM models). Because the $\Phi b\bar{b}$ coupling varies as,  $g_{\Phi b\bar{b}}\sim 1/v_\chi$ in the eGM model ($g_{\Phi b\bar{b}}\sim 1/v_\Delta$ in the GM model). In the mass range where $m_h \lesssim 2m_\Phi - \Gamma_\Phi$, the partial decay width for the above three-body decays is given in~\cite{Djouadi:1995gv}. Additionally, in the presence of a light singly-charged scalar $H^+_3$ in the GM model, a top quark decay via $t \rightarrow H^+_3 b$ can alter the branching fractions of final states containing SM particles. Consequently, these observables are crucial for constraining the parameter space of the eGM model, as both singly-charged scalars $(H^+,F^+)$ couple to fermions. In our analysis, we include the latest ATLAS and CMS upper limits on $\mathcal{B}(t\to H^+b)$ and $\mathcal{B}(t\to F^+b)$ for the eGM model, and $\mathcal{B}(t\to H^+_3b)$ for the GM model, where the charged scalars subsequently decay into one of the following channels: $\tau^+\nu$,  $c\bar{s}$, and $c\bar{b}$, as reported in Tables~\ref{tab:CMS_charged} and~\ref{tab:ATLAS_charged}. In addition to that, we include the latest LHC and LEP data on left-handed slepton searches, $pp\,(e^+e^-) \to \tilde{l}_L^+ \tilde{l}_L^-\to l^+l^- +2\tilde{\chi}^0$, which can be reinterpreted as  $pp\,(e^+e^-) \to H^+ H^-\to \tau^+\tau^- +2\nu$ or $pp\,(e^+e^-) \to F^+ F^-\to \tau^+\tau^- +2\nu$ in the eGM model, and $pp\,(e^+e^-) \to H_3^+ H_3^-\to \tau^+\tau^- +2\nu$ in the GM model, as listed in Table~\ref{tab:direct_search}. In the presence of a light neutral scalar $m_\Phi=95$ GeV, the relative branching fractions of the singly-charged scalars decaying into $\tau^+ \nu$ are reduced significantly 
for both models. This is because, in the low-mass range, the three-body decays of the singly-charged scalars into $\Phi W^{+*}$ have sizable branching ratios ($\Phi=H$ in both the  GM and eGM models).

The ATLAS and CMS searches for light doubly-charged scalars in the leptonic final state assume a 100\% branching fraction (see Refs.~\cite{CMS:2012dun,ATLAS:2014kca,CMS:2016cpz,CMS:2017pet}), since in this case the decay to vector boson final states is suppressed.  In the GM (eGM) model, the process, $H_5^{++} \rightarrow \ell^+ \ell^+$  ($F^{++} \rightarrow \ell^+ \ell^+$) proceeds exclusively via the lepton number-violating interaction involving the lepton doublet and the Higgs triplet fields. As discussed in Section~\ref{sec:model}, this interaction is negligible for $v_\chi \gg 10^{-4}$~GeV, where the doubly-charged scalar decays predominantly into $W^{+}W^{+}$ and the leptonic branching fraction is vanishingly small. Therefore, we do not consider these searches involving a light doubly-charged scalar in our study. We incorporate searches for light and heavy neutral scalars decaying into two photons or two vector bosons in both models, as summarised in Tables~\ref{tab:ATLAS_light}, \ref{tab:CMS_light}, \ref{tab:CMS_heavy_scalar_to_bosons}, and~\ref{tab:ATLAS_heavy_scalar_to_bosons}. We include searches for a light pseudoscalar state $A$ in the eGM model ($H_3^0$ in the GM model) decaying into a pair of leptons, as reported by the ATLAS and CMS collaborations, given in Tables~\ref{tab:ATLAS_light} and~\ref{tab:CMS_light}. In the GM model, the pseudoscalar $H^0_3$ and the singly-charged scalars $H^\pm_3$ are mass-degenerate. Consequently, the lower bound on $m_{H^0_3}$ is set by the LEP limit on $m_{H^\pm_3}$, i.e. $m_3= 80$ GeV. All theoretical and experimental datasets for $\sigma\cdot\mathcal {B}$ used in this work are discrete and are linearly interpolated throughout the analysis.  

\section{Results}\label{sec:results}

In this section, we present the allowed parameter space, taking into account all theoretical constraints and the 95.4\% probability regions from the combined fits to experimental data for both the GM and eGM models. For the theoretical constraints, we assume flat likelihoods, and the allowed regions represent 100\% probability contours. The results are based on the prior ranges specified in Table~\ref{tab:GMeGMprior} for the GM and eGM models, unless stated otherwise. 

\begin{figure}[!h]
     \centering
             \includegraphics[clip=true,width=0.85\columnwidth]{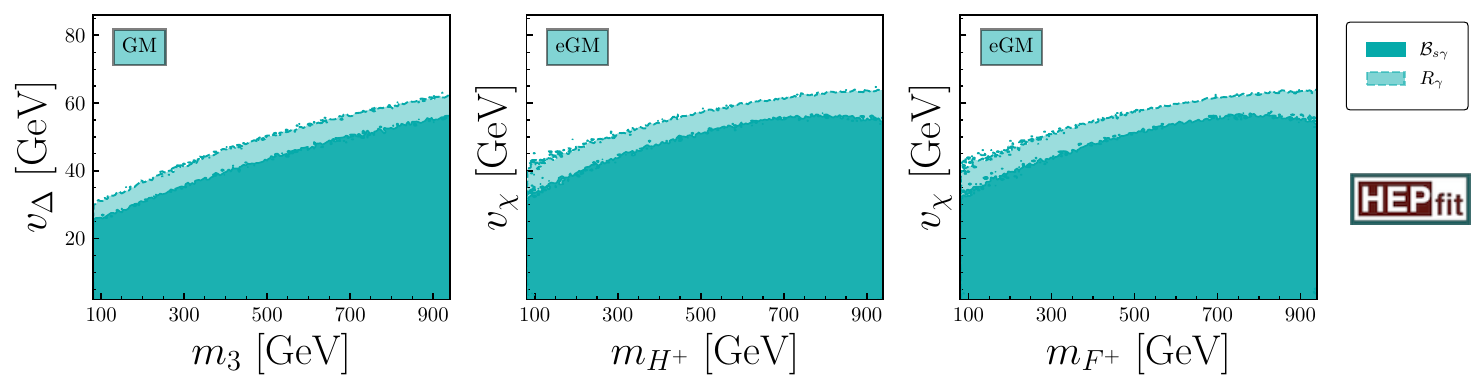}
     \caption{Left: Constraints on  $m_3$ vs. $v_\Delta$ plane from $R_\gamma$ and $\mathcal{B}_{s \gamma}$ for the GM model. Middle (Right): Constraints on $m_{H^+}$ vs. $v_\chi$ ($m_{F^+}$ vs. $v_\chi$) plane from $R_\gamma$ and $\mathcal{B}_{s \gamma}$ for the eGM model. The dark cyan regions are allowed from $\mathcal{B}_{s \gamma}$, and the light cyan regions are allowed from $R_\gamma$ constraints at a 95.4\% probability.}
     \label{fig:2}
\end{figure}

In Figure~\ref{fig:2}, we examine the impact of the $B$-physics observables, $\mathcal{B}_{s \gamma}$ and $R_\gamma$, on the parameter space of both the GM and eGM models.\footnote{For this, we sample the parameter space over the range,
$v_\Delta\in[0,86]$ GeV for the GM model, $v_\chi\in[0,86]$ GeV for the eGM model, and the BSM scalar masses, $m_i\in[0.08,1]$ TeV for both models. The remaining priors for the GM and eGM models are specified in Table~\ref{tab:GMeGMprior}.} In the GM model, only the singly-charged scalar from the custodial triplet contributes to these $B$-physics observables. Each of these observables imposes an upper bound on the triplet VEV, $v_\Delta$, as a function of the singly-charged scalar mass, $m_3$, as shown in the left panel of Figure~\ref{fig:2}. Whereas in the eGM model, both singly-charged scalars, $H^+$ and $F^+$, contribute to these $B$-physics observables. The middle (right) panel of Figure~\ref{fig:2} presents the corresponding allowed parameter space in the $m_{H^+}$ vs. $v_\chi$ ($m_{F^+}$ vs. $v_\chi$) plane for the eGM model. In both models, we observe that the latest result from HFLAV~\cite{HFLAV:2022esi} on $\mathcal{B}_{s \gamma}$ imposes the dominant constraint on the triplet VEV. Therefore, from this point onward, we consider only $\mathcal{B}_{s \gamma}$ when discussing $B$-physics constraints. As seen in Figure~\ref{fig:2}, the triplet VEV is tightly constrained in the lower mass region of the singly-charged scalars. This behaviour originates from the singly-charged scalar contributions to the Wilson coefficients $\Delta C_7$ and $\Delta C_8$, which become sizeable for light singly-charged scalars and large $\tan\beta$ (see Eq.~(\ref{eq:deltac78})). Consequently, smaller values of the triplet VEV are required to keep $\mathcal{B}(\bar{B}\to X_s\gamma)$ within the experimentally allowed range given in Eq.~(\ref{eq:flavor_expt}). This demonstrates that the $B$-physics constraints are particularly important in the lower mass region of the singly-charged scalars.

In the GM model, the inclusion of the latest limit on $\mathcal{B}_{s \gamma}$ from HFLAV~\cite{HFLAV:2022esi} slightly tightens the upper bound on $v_\Delta$ compared to the results presented in Ref.~\cite{Hartling:2014aga}. We further see that the upper bound on  $v_\chi$ in the eGM model is less stringent for low $m_{H^+}$ or  $m_{F^+}$ compared to the corresponding upper bound on $v_\Delta$ for low $m_3$ in the GM model. However, for large $m_{H^+}$ or $m_{F^+}$, the upper bounds on the triplet VEVs become comparable in both models. This behaviour arises because, at low $m_{H^+}$ or $m_{F^+}$, the contribution to $\Delta C_{7,8}$ in the eGM model is maximised and reproduces the GM
result when $m_{H^+}$ and $m_{F^+}$ are degenerate. As the mass splitting between $m_{H^+}$ and $m_{F^+}$ increases, the heavier state contributes less significantly to $\Delta C_{7}$ and $\Delta C_{8}$, resulting in weaker constraints on $v_\chi$. In the limit where both $m_{H^+}$ and $m_{F^+}$ are heavy, the contributions to $\Delta C_7$ and $\Delta C_8$ saturate and become nearly identical in the two models, leading to comparable upper bounds on the triplet VEVs.

\begin{figure}[!h]
     \centering
             \includegraphics[clip=true,width=0.7\columnwidth]{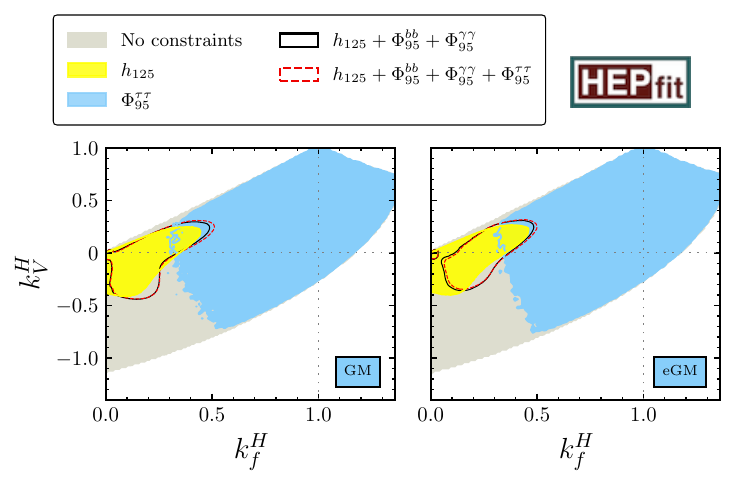}
     \caption{Left: $\kappa^H_f$ vs. $\kappa^H_V$ plane for the GM model. Right: $\kappa^H_f$ vs. $\kappa^H_V$ plane for the eGM model. The grey regions are the allowed parameter space for a light Higgs $H$ without imposing any constraints, where $m_H<m_h\approx 125$ GeV. The yellow regions are consistent with the signal strengths of the SM-like Higgs boson $h$ (denoted $h_{125}$). The light blue regions are consistent with the CMS di-tau excess at 95.4~GeV (where $\Phi_{95}$ is identified with $H$). The dashed red and solid black contours show the regions allowed by the global fit with and without the $\tau^+\tau^-$ channel, respectively.}
     \label{fig:4}
\end{figure}

Next, we discuss the local excesses near 95~GeV reported at the LHC~\cite{CMS:2024yhz,CMS-HIG-20-002,ATLAS:2024bjr} and at LEP~\cite{LEPWorkingGroupforHiggsbosonsearches:2003ing}. We examine whether a CP-even scalar $H$ with a mass of 95.4~GeV can account for these excesses within the CP-conserving GM and eGM models. The grey regions in Figure~\ref{fig:4} show the prior range in the $\kappa^H_f$--$\kappa^H_V$ plane for a light Higgs boson $H$ with $m_H < m_h \approx 125$~GeV.\footnote{In contrast to the case with $m_{H} > m_h$, where the mixing angle $\alpha < 0$ is generally favoured (see Refs.~\cite{Chiang:2018cgb,Chowdhury:2024mfu}), our analysis focuses on the scenario where $m_{H} < m_h$, for which positive values of $\alpha$ are typically preferred. We therefore sample in the region $\alpha\in [0,\pi/2]$.} The yellow regions correspond to the parameter space allowed by the combined signal strength measurements of the SM-like Higgs boson $h$ decaying into all final states, as summarised in Tables~\ref{tab:signalstrengthsAC78}, \ref{tab:signalstrengthsC13}, and \ref{tab:signalstrengthsA13}. The light blue regions are consistent with the signal strength $\mu^{\tau\tau}_{\text{pp}}$ of the CMS di-tau excess (see Table~\ref{tab:95 signal strength}), interpreted as $H \to \tau^+\tau^-$ with $m_H = 95.4$~GeV. The dashed red and solid black contours show the regions allowed by the global fit with and without the $\tau^+\tau^-$ channel, respectively. 

As seen in Figure~\ref{fig:4}, neither the CP-conserving GM nor the eGM model can reproduce the central value of the CMS di-tau excess. This tension can be understood analytically. The measured signal strength, $\mu^{\tau\tau}_{pp} = 1.23 \pm 0.55$, favours nearly SM-like Yukawa couplings for the light scalar, $|\kappa_f^H| \simeq 1$. In the GM and eGM models, the couplings of both custodial singlets are determined by the same two angles, $\kappa_f^h = s_\alpha/c_\beta$ and $\kappa_f^H = c_\alpha/c_\beta$. A large $\kappa_f^H$ therefore requires a suppressed $\kappa_f^h$, which is disfavoured by the $h$ signal strength data (see the yellow regions in Figure~\ref{fig:4}). Nevertheless, both models remain compatible with the CMS di-tau measurement at the $\sim 2\sigma$ level. Within the Bayesian framework, we quantify the impact of this tension by computing the Information Criterion (IC)~\cite{Ando2007,Ando01012011} for the GM and eGM models, with and without the CMS di-tau excess included in the global fit. The IC is computed as
\begin{equation}
  \mathrm{IC} = -2\,\overline{\ln \mathcal{L}} + 4\,\sigma^{2}_{\ln \mathcal{L}} ,
  \label{eq:IC}
\end{equation}
where $\overline{\ln \mathcal{L}}$ and $\sigma^{2}_{\ln \mathcal{L}}$ are the mean and variance of the log-likelihood. The first term reflects the average quality of the fit, while the second penalises the effective number of parameters constrained by the data. The IC values are summarised in Table~\ref{tab:IC_tautau}.
\begin{table}[!ht]
\begin{center}
\setlength{\tabcolsep}{5pt}
\renewcommand{\arraystretch}{1.3}
\scalebox{0.85}{
\begin{tabular}{| c| c|c|}
\hline
\;\;Observable\;\; & \;\; IC$_{\rm GM}$ \;\;& \;\;IC$_{\rm eGM}$\;\;\\
\hline\hline
 $h_{125}$ + $\Phi_{95}^{\gamma\gamma}$ + $\Phi_{95}^{b b}$ & \;\;48.466\;\; & \;\;39.058 \;\; \\[2pt]
 \hline
 $h_{125}$ + $\Phi_{95}^{\gamma\gamma}$ + $\Phi_{95}^{b b}+\Phi_{95}^{\tau\tau}$ & \;\;50.198 \;\;& \;\;39.198 \;\;\\[2pt]
\hline
\end{tabular}}
\caption{Information Criterion for the GM/eGM model, with and without the CMS di-tau measurement in the likelihood.}
\label{tab:IC_tautau}
\end{center}
\end{table}
Including the di-tau channel increases the IC of both models, while its impact on the allowed parameter space remains mild (see red dashed contour in Figure~\ref{fig:4}). In our combined global fit for both the GM and eGM models, we include only the $\gamma\gamma$ and $b\bar b$ excesses at 95~GeV. The di-tau excess has so far been reported only by CMS, with its largest local significance at a mass closer to 100~GeV. Moreover, a CP-even interpretation of this excess is disfavoured by the absence of a corresponding signal in the CMS searches for $t\bar{t}$-associated production with di-tau decays based on the full Run~2 dataset~\cite{CMS:2024ulc}, as shown in Ref.~\cite{Biekotter:2023jld}.

\begin{figure}[!h]
     \centering
             \includegraphics[clip=true,width=0.8\columnwidth]{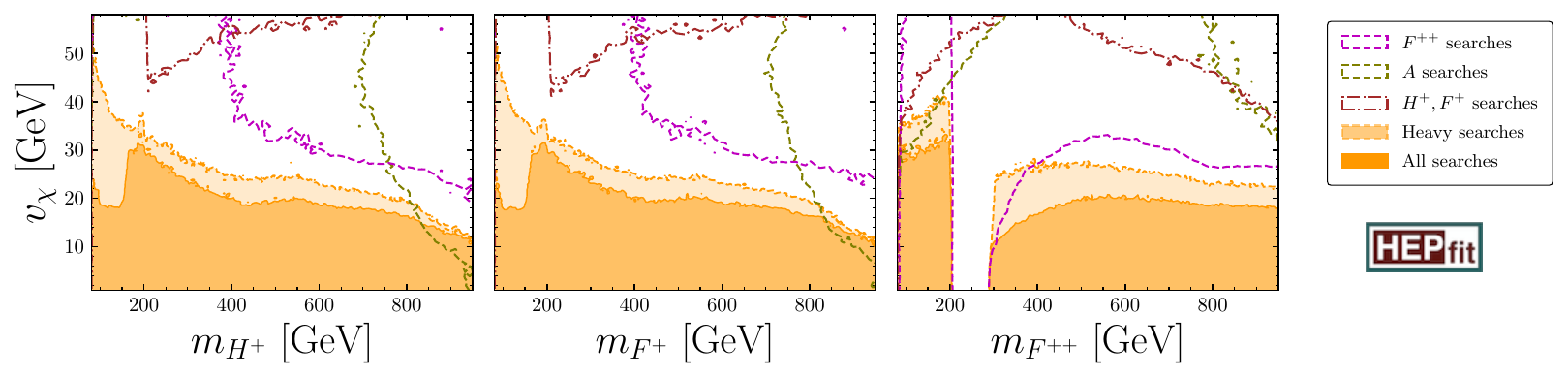}
     \caption{Impacts of different direct search limits on $v_\chi$ vs. $m_{H^+}$, $m_{F^+}$, and $m_{F^{++}}$ planes at a 95.4\% probability. The regions above the magenta dashed, olive green dashed, and maroon dot-dashed lines are excluded from $F^{++}$, $A$, $H^+$ (and $F^+$) searches, respectively. The light orange regions are allowed by heavy scalar searches, and the dark orange areas indicate regions allowed from the combined fit to both light and heavy scalar direct search data. The direct search limits on the singly and doubly-charged scalars considered in this analysis are listed in Table~\ref{tab:direct_search}.}
     \label{fig:3}
\end{figure}

\begin{figure}[!h]
     \centering
             \includegraphics[clip=true,width=0.8\columnwidth]{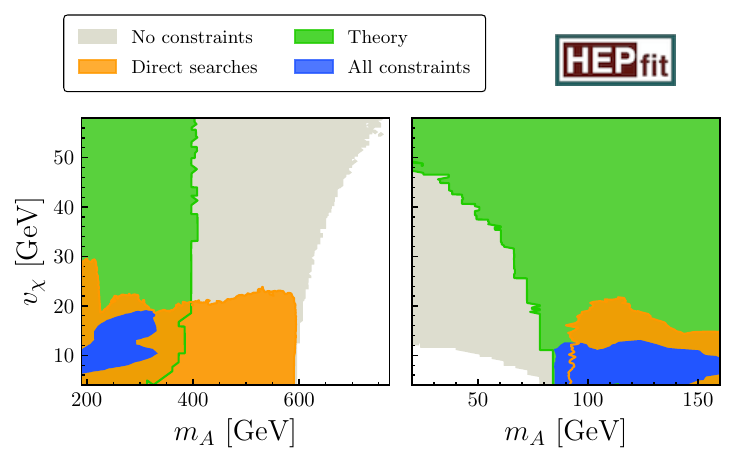}
     \caption{Allowed regions in the $m_A$ vs. $v_\chi$ plane for the eGM model. The grey regions denote the allowed region in the presence of a light Higgs $H$ with $m_H\in[90,100]$ GeV without imposing any constraints. The green regions show the parameter space allowed by the theoretical constraints. The orange regions represent the parameter space consistent with direct search constraints on all BSM scalars in the eGM model. The blue regions represent the parameter space allowed by a combined fit to the 95 GeV Higgs signal strengths in the $\gamma \gamma $ and $b\bar{b}$ channels, along with all the experimental and theoretical constraints.}
     \label{fig:5}
\end{figure}

In Figure~\ref{fig:3}, we present the constraints on $v_\chi$ as a function of $m_{H^+}$, $m_{F^+}$, and $m_{F^{++}}$ in the eGM model, derived from the LHC direct search limits. This result is based on the prior range of the eGM model given in Table~\ref{tab:GMeGMprior}. Since we restrict $m_H$ in the range $m_H\in [90,100]$~GeV, direct search limits for $H$ are not considered in our analysis. The light orange regions in Figure~\ref{fig:3} represent the parameter space allowed by heavy scalar searches, while the dark orange areas are allowed by the combined constraints from both light and heavy scalar searches at a 95.4\% probability. In the left and middle panels, the parameter space of $m_{H^{+}}, m_{F^{+}} < 200$~GeV is constrained by top quark decays, $t \rightarrow H^+ b$ and $t \rightarrow F^+ b$. For $m_{H^{+}}, m_{F^{+}} > 200$ GeV, direct searches for all heavy BSM scalars constrain the triplet VEV, $v_\chi$. When $m_{H^{+}}, m_{F^{+}} > 700$ GeV, the decay channels $H^+, F^+ \rightarrow W^+ A$ provide the most stringent constraints on $v_\chi$, shown in olive dashed lines in Figure~\ref{fig:3}. This is because the pseudoscalar masses, $m_A>600$ GeV, are nearly excluded with a 95.4\% probability from direct search limits, as shown in Figure~\ref{fig:5}.\footnote{Note that the pseudoscalar mass $m_A$ is not an independent parameter in the eGM model (see Eq.~(10) of Ref.~\cite{Chowdhury:2024mfu}).} In the right panel, the strongest constraint on the $m_{F^{++}}$ vs. $v_\chi$ plane for $m_{F^{++}} > 200$ GeV comes from the direct search limits of $F^{++}$, shown in magenta dashed lines in Figure~\ref{fig:3}.\footnote{The region where $m_{F^{++}}\sim\!450$ GeV is less constrained due to a recent excess reported by ATLAS near 450 GeV~\cite{ATLAS:2024txt} in the same-sign $WW$ channel.} The $F^{++}$ boson masses in the range 200 GeV to $\sim 300$ GeV are predominantly excluded by the pair production mode of $F^{++}$ bosons~\cite{ATLAS:2021jol}. In contrast, the region with $m_{F^{++}} < 200$ GeV remains unconstrained by direct searches for $F^{++}$ alone.\footnote{Ref.~\cite{Ashanujjaman:2022ofg} provides a detailed phenomenological study of the doubly-charged boson in the type-II seesaw model within the mass range of 84--200 GeV.} However, this low-mass region becomes constrained once searches for the light pseudoscalar $A$ and the light singly-charged scalars $F^+$ and $H^+$ are also included. 

Figure~\ref{fig:5} presents the allowed range of $v_\chi$ as a function of the pseudoscalar mass $m_A$ in the eGM model.\footnote{In the GM model, the pseudoscalar and one of the singly-charged scalar masses are degenerate, so the lower bound on the pseudoscalar mass is determined by the LEP constraint on the charged scalar, which is around 80 GeV. In this work, we investigate how light the pseudoscalar $A$ can be within the framework of the CP-conserving eGM model.} The left panel displays the higher $m_A$ mass range, while the right panel focuses on the lower mass region, $m_A\in[10,180]$ GeV. In the CP-conserving eGM model, the masses and mixing angles of singly-charged scalars $H^+$ and $F^+$, together with the triplet VEV $v_\chi$, determine the pseudoscalar mass $m_A$~\cite{Chowdhury:2024mfu}. As a result, the entire $m_A-v_\chi$ parameter space is not accessible. The grey regions denote the allowed parameter space for an additional light scalar with $m_H\in[90,100]$~GeV without imposing any theoretical or experimental constraints. There exists an upper bound on $m_A$ for any fixed value of $v_\chi$, and a lower bound on $v_\chi$ for fixed $m_A$ in the range $m_A\in[10,80]$ GeV, both arising solely from the model properties and parameter sampling. Imposing the
theoretical constraints, such as NLO unitarity and BFB conditions,
further reduces the accessible parameter space, shown by the green regions. The orange regions represent the parameter space consistent with direct search limits on all BSM scalars in the eGM model. In the low-mass region, the direct search limit imposes a strong upper bound on $v_\chi$, excluding most of the parameter space below approximately 80 GeV. In the range $m_A\in[80,180]$ GeV, direct search limits impose an upper bound on $v_\chi$ of around 22 GeV, while for $m_A>180$ GeV, the upper bound increases to $\sim 30$ GeV. The blue regions indicate the parameter space compatible with a combined fit to the 95 GeV Higgs  signal strength data, along with all other experimental and theoretical constraints. Here, the 95 GeV signal strength data includes the $\gamma\gamma$ and $b\bar{b}$ decay channels, given in Table~\ref{tab:95 signal strength}. This combined fit further tightens the upper bounds on the triplet VEV, reducing it to approximately 12 GeV for $m_A\in[80,180]$ GeV, and to about 20 GeV for $m_A>180$ GeV.

\begin{figure}[!h]
     \centering
             \includegraphics[clip=true,width=0.8\columnwidth]{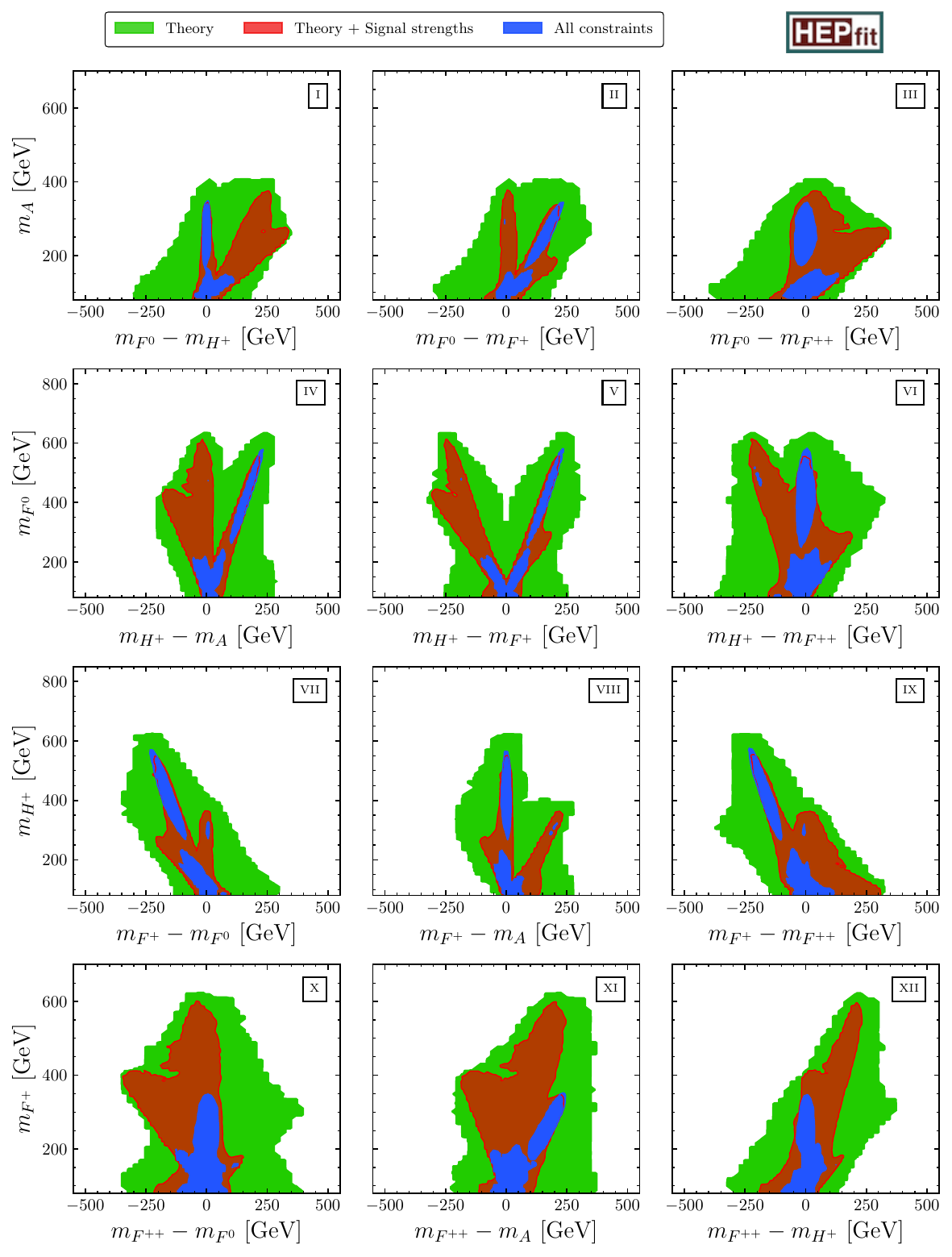}
     \caption{Allowed regions in the $m_i$ vs. $(m_j-m_k)$ planes in the eGM model. The red regions show the parameter space allowed from a fit to the combined Higgs signal strength data, including theory bounds. The green and blue regions have the same meaning as in Figure~\ref{fig:5}.}
     \label{fig:6}
\end{figure}

In Figure~\ref{fig:6}, we present the allowed ranges for the BSM scalar masses and their mass differences in the eGM model, after applying theoretical and experimental constraints. The green regions correspond to parameter space allowed by the theoretical bounds alone. A comprehensive analysis of the impact of these theoretical constraints can be found in Ref.~\cite{Chowdhury:2024mfu}. The presence of a light BSM scalar $H$ with $m_H \in [90,100]$ GeV imposes upper bounds on the other BSM scalar masses and their mass differences, as reflected in the green region of Figure~\ref{fig:6}. In addition, imposing the combined Higgs signal strength data further restricts the triplet vacuum expectation value $v_\chi$ and the mixing angle $\alpha$~\cite{Chowdhury:2024mfu}, reducing the allowed region to the red area.\footnote{In this work, combined Higgs signal strength data includes both 125 GeV and 95 GeV Higgs signal strength data from the LHC and LEP colliders.}  Finally, the blue regions represent the parameter space that remains consistent with
all theoretical and experimental constraints, providing the strongest bounds on the BSM scalar
masses and their mass differences. The maximum allowed mass splitting among the BSM scalars is approximately 250 GeV. In the limit where $m_A = m_{H^+}$ and $m_{F^0} = m_{F^+} = m_{F^{++}}$ with the mixing angle $\delta=0$, the eGM model reduces to the GM model (see Section~\ref{sec:model}).
In our fit, the masses of the two singly-charged scalars, $m_{H^\pm}$ and
$m_{F^\pm}$, are sampled independently. In the calculation, the lighter eigenstate is identified with
$H^\pm$ and the heavier one with $F^\pm$, while Figure~\ref{fig:6} shows the posteriors in
terms of the sampled masses. In Figure~\ref{fig:6}, the bands with $m_{F^0} \simeq m_{H^\pm}$ (panel~I), $m_{H^\pm} \simeq m_{F^{\pm\pm}}$ (panels~VI and XII) and $m_{F^\pm} \simeq m_A$ (panel~VIII) correspond to the GM limit.\footnote{In the GM limit, one of the two singly charged eigenstates coincides with the singly charged member of the GM fiveplet, $H_5^\pm$, and becomes nearly degenerate with $F^{\pm\pm}$ and $F^0$, while the other is identical to  $H_3^\pm$.}  Similarly, the band with $m_{F^0} \simeq m_{F^{\pm\pm}}$ in panels III and X is also a consequence of the GM limit. All the mass planes in the eGM model are displayed in Figure~\ref{fig:9} in Appendix~\ref{app:C}. Combining all constraints, the $95.4\%$ probability upper limits on the eGM model parameters and the allowed mass splittings are given in Table~\ref{tab:eGM_limits}.

\begin{table}[!h]
\begin{center}
\begin{tabular}{| l ll|lll|}
\hline
  {\;\small Parameters}\;\;&  &{\;\;\small Upper limit (in GeV)  }  & {\small \;Parameters\;} &  & {\;\;\small Upper limit (in GeV)  } \\
\hline
\;\;$v_\chi$ &: & \;\; $\lesssim 20$  & \;\;$m_{F^{++}}$ &\;\;: & \;\;$\lesssim 600 $\; \\
\;\;$m_{H^{+}}$  &: &  \;\;$\lesssim 590$ & \;\;$m_{F^{+}}$  &\;\;:& \;\; $\lesssim 500$ \\
\;\;$m_{A}$  &: &  \;\;$\lesssim 350$ & \;\;$m_{F^{0}}$  &\;\;:& \;\; $\lesssim 580$ \\
\;\;$|m_{H^{+}}-m_{A}|$  &: &  \;\;$\lesssim 240$ & \;\;$|m_{F^{+}}-m_{A}|$  &\;\;:& \;\; $\lesssim 210$ \\
\;\;$|m_{H^{+}}-m_{F^+}|$  &: &  \;\;$\lesssim 250$ & \;\;$|m_{F^{+}}-m_{F^0}|$  &\;\;:& \;\; $\lesssim 250$ \\
\;\;$|m_{H^{+}}-m_{F^{++}}|$  &: &  \;\;$\lesssim 220$ & \;\;$|m_{F^{+}}-m_{F^{++}}|$  &\;\;:& \;\; $\lesssim 240$ \\
\;\;$|m_{F^{0}}-m_{H^+}|$  &: &  \;\;$\lesssim 110$ & \;\;$|m_{F^{++}}-m_{F^0}|$  &\;\;:& \;\; $\lesssim 140$ \\
 \;\;$|m_{F^{++}}-m_{A}|$  &:& \;\; $\lesssim 250$ &&&\\

\hline
\end{tabular}
\caption{$95.4\%$ probability upper limits on the eGM model parameters and the allowed mass splittings.}
\label{tab:eGM_limits}
\end{center}
\end{table}

\begin{figure}[!h]
     \centering
             \includegraphics[clip=true,width=0.8\columnwidth]{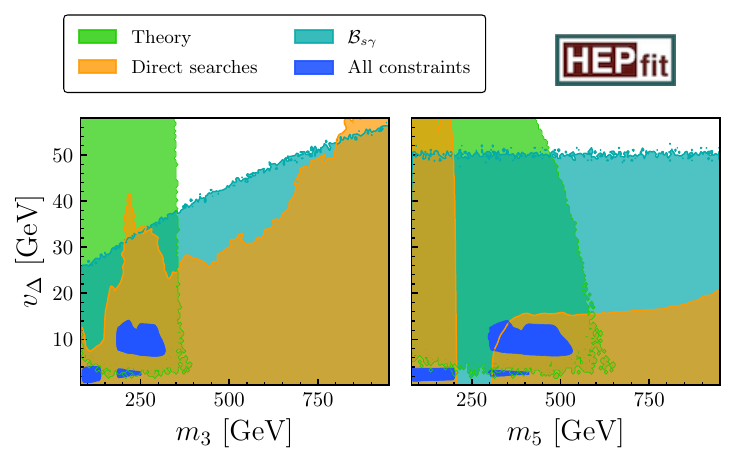}
     \caption{Allowed regions in the $m_3$ vs. $v_\Delta$ plane (left) and $m_5$ vs. $v_\Delta$ plane (right) for the GM model. The green, cyan, orange, and blue regions have the same meaning as in Figures~\ref{fig:2} and~\ref{fig:5}.}
     \label{fig:7}
\end{figure}

\begin{figure}[!h]
     \centering
             \includegraphics[clip=true,width=0.75\columnwidth]{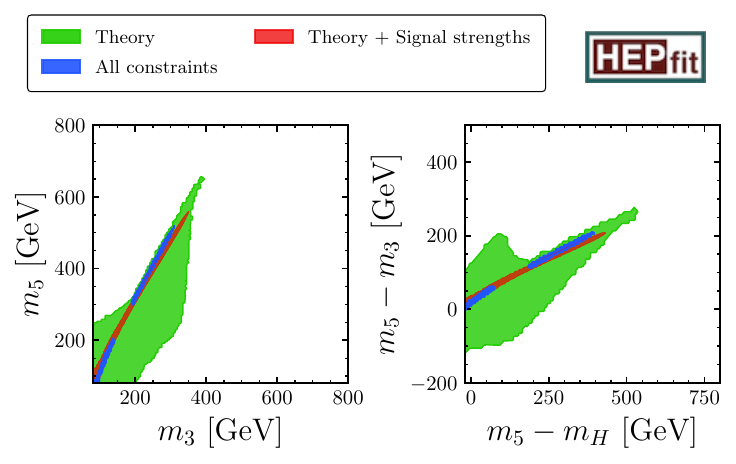}
     \caption{Allowed regions in the mass (left) and mass difference (right) planes in the GM model. The green, red, and blue regions have the same meaning as in Figure~\ref{fig:6}.}
     \label{fig:8}
\end{figure}

In Figure~\ref{fig:7}, we present the constraints on the triplet VEV $v_\Delta$ as a function of BSM scalar masses in the GM model. The figure incorporates theoretical constraints, direct search limits, $B$-physics observables, and a combined fit to the 95 GeV Higgs signal strength data, along with all other relevant experimental and theoretical constraints.\footnote{We sample the parameter space over the BSM scalar masses, $m_{H},m_3,m_5\in[0.08,1]$ TeV when presenting the allowed parameter space from $B$-physics and direct search limits. The remaining priors for the GM model are given in Table~\ref{tab:GMeGMprior}.} The colour scheme follows that of Figures~\ref{fig:2} and \ref{fig:5}. As discussed in Section~\ref{sec:fitting set-up and constraints}, stringent LEP bounds on low-mass charged scalar necessitate a lower limit of 80 GeV on both $m_3$ and $m_5$ in our analysis. Theoretical constraints place an upper bound on these masses for a fixed $v_\Delta$, while $B$-physics data limit $v_\Delta$ for a fixed $m_3$, as illustrated in Figure~\ref{fig:2}. In the left panel, the low $m_3$ region is constrained by bounds from $t \rightarrow H^+_3 b$ decays, while the region with $m_3 > 180$ GeV is dominantly constrained by searches for the pseudoscalar $H^0_3$ and the singly-charged scalar $H_3^+$. In the right panel, the region where $m_5 < 200$ GeV remains unconstrained by direct searches, whereas for $m_5 > 200$ GeV, the parameter space is significantly restricted by all heavy scalar searches. In particular, the mass range $200~\mathrm{GeV}\lesssim m_5\lesssim300~\mathrm{GeV}$ is predominantly excluded by pair production of $H_5^{++}$~\cite{ATLAS:2021jol}. In the combined fit, the mass difference between $m_5$ and $m_3$ is tightly constrained (see Figure~\ref{fig:8}). 
Consequently, the constraints on one mass plane can be directly mapped onto the other mass plane. For example, the stringent limits on $v_\Delta$ at low $m_3$ indirectly constrain the low-$m_5$ region, while the strong direct search limits at high $m_5$ substantially reduce the allowed values of $v_\Delta$ for large $m_3$. Overall, the combined fit restricts the triplet VEV to approximately $v_\Delta\lesssim4$~GeV for $m_3\in[80,140]$~GeV and $m_5\in[80,200]$~GeV. For larger scalar masses, namely $m_3\in[190,320]$~GeV and $m_5\in[300,530]$~GeV, the upper bound on the triplet VEV relaxes to about $15$~GeV.

Finally, in Figure~\ref{fig:8}, we present the allowed parameter space for the BSM scalar masses and their mass differences within the GM model. The colour scheme is consistent with that used in Figure~\ref{fig:6}. For a comprehensive discussion of the impact of theoretical constraints on the GM model's parameter space, we refer the reader to Ref.~\cite{Chowdhury:2024mfu}. Fixing $m_{H} \in [90,100]$~GeV imposes upper bounds on the masses and mass differences of the other BSM scalars, as shown in the green region of Figure~\ref{fig:8}. The theoretical constraints alone allow both positive and negative values of the mass difference between $m_5$ and $m_3$. However, the combined Higgs signal strength data further restrict the triplet VEV $v_\Delta$ and the mixing angle $\alpha$~\cite{Chowdhury:2024mfu}, thereby reducing allowed parameter space to the red region, and leading to a preferred mass hierarchy between $m_3$ and $m_5$. This is because, in the GM model, the two trilinear parameters $\tilde\mu_1$ and $\tilde\mu_2$ are not constrained by theory bounds. Fixing $m_{H} \in [90,100]$ GeV uniquely determines one of these parameters. The other one is constrained by the Higgs signal strength data along with the triplet VEV $v_\Delta$, and the mixing angle $\alpha$. As a result, the combined fit, including both theoretical constraints and signal strength data, leads to the mass hierarchy $m_5>m_3$ for $m_5,m_3\gtrsim 100$ GeV. This behaviour can be understood from the tree-level
mass relations. In the limit of small $\alpha$ and $v_\Delta$,
Eqs.~(\ref{eq:m5m3GM}) and (\ref{eq:mH1GM}) imply $m_5^2 - m_3^2 \simeq 2\left(m_3^2 - m_{H}^2\right).$ Hence, $m_3$ lies between $m_{H}$ and $m_5$, so a light $H$
necessarily implies $m_5 > m_3$ in the high-mass region. The blue areas in Figure~\ref{fig:8} are allowed from the combined fit to all the theoretical and experimental constraints. Combining all constraints, the $95.4\%$ probability upper limits on the GM model parameters and the allowed mass splittings are given in Table~\ref{tab:GM_limits}.
\begin{table}[!h]
\begin{center}
\begin{tabular}{| l ll|lll|}
\hline
  {\;\small Parameters}\;\;&  &{\;\;\small Upper limit (in GeV)  }  & {\small \;Parameters\;} &  & {\;\;\small Upper limit (in GeV)  } \\
\hline
\;\;$v_\Delta$ &: & \;\; $\lesssim 15$  & \;\;$m_5$ &\;\;: & \;\;$\lesssim 530 $\; \\
\;\;$m_3$  &: &  \;\;$\lesssim 320$ & \;\;$|m_5-m_3|$  &\;\;:& \;\; $\lesssim 210$ \\
\hline
\end{tabular}
\caption{$95.4\%$ probability upper limits on the GM model parameters and the allowed mass splittings.}
\label{tab:GM_limits}
\end{center}
\end{table}

Furthermore, the preferred mass hierarchy $m_5 > m_3$ kinematically forbids the decays
$H_3^0 \to H_5^0 Z$, $H_3^+ \to H_5^+ Z$,
$H_3^+ \to H_5^0 W^+$,
$H_3^+ \to H_5^{++} W^-$. These channels are phenomenologically relevant in the low-$m_5$ and low-$v_\Delta$ region of the GM model parameter space~\cite{Logan:2017jpr}.

\section{Summary and Outlook}\label{sec:conclusion}
We have performed a comprehensive Bayesian global fit of the GM and eGM models in the presence of an additional CP-even scalar with a mass around 95 GeV. The global fit combines the latest Higgs signal strength measurements for the 95 GeV and 125 GeV Higgs bosons, direct searches for light and heavy scalars at the LHC and LEP, $B$-physics observables, and improved theoretical constraints from NLO unitarity and BFB conditions.

Our analysis shows that the LEP $b\bar b$ and LHC di-photon excesses around 95 GeV can be accommodated consistently with the observed properties of the 125 GeV scalar at the LHC. For the CMS di-tau excess, neither the CP-conserving
GM nor the eGM model can reproduce the central value of the measurement. However, both models remain compatible with it at the $\sim 2\sigma$ level. We further show that an additional 95 GeV Higgs boson leads to a relatively light scalar spectrum in the GM and eGM models. We obtain the following upper bounds on the model parameters, with a 95.4\% probability:
\begin{enumerate}
\renewcommand{\labelenumi}{(\roman{enumi})}
    \item in the eGM model, the triplet VEV is constrained to
be approximately below 20 GeV, and the remaining BSM scalar masses are constrained to $m_{F^{++}} \lesssim 600$ GeV, $m_{F^{+}} \lesssim 500$ GeV, $m_{F^{0}} \lesssim 580$ GeV, $m_{H^{+}} \lesssim 590$ GeV, and $m_{A} \lesssim 350$ GeV, and their mass splitting is bounded to be below approximately 250 GeV;
\item in
the GM model, the triplet VEV is limited to below 15 GeV, the custodial quintet and triplet scalar masses are constrained to be below 530 GeV and 320 GeV, respectively, with the mass hierarchy $m_5>m_3$ when $m_5,m_3\gtrsim 100$ GeV.
\end{enumerate}
The results presented in this work demonstrate that interpreting the 95 GeV excesses as an additional light Higgs boson makes both GM and eGM models significantly more predictive, with the GM model being subject to more stringent constraints than the eGM model.

\subsection{Outlook}

An important consequence of the preferred mass hierarchy $m_5>m_3$ in the GM model is that the cascade decays $H_5\to H_3V$ and  $H_5\to H_3H_3$ can become dominant. These channels suppress the branching fractions of the  $H_5\to VV$ modes, thereby reducing the sensitivity of the existing $H_5$ searches at the LHC based on di-boson final states. Thus, our results motivate dedicated searches for cascade decays beyond the conventional $VV$ channels~\cite{Logan:2017jpr,Logan:2018wtm,Ismail:2020kqz,Cacciapaglia:2022bax,Banerjee:2025qkb}. The projected sensitivity of high luminosity-LHC (HL-LHC) VBF searches is expected to probe this region more effectively and place more stringent constraints on the triplet VEV.

Additionally, the richer scalar sector of the eGM model yields qualitatively different collider signatures compared
to the GM model, motivating dedicated searches at the HL-LHC and future colliders.
\begin{itemize}
    \item In the eGM model, additional cascade decays, such as $F^{++}\to F^+W^+$, $F^+H^+$, and $F^+F^+$, suppress the $F^{++}\to W^+W^+$ signature, leading to weaker constraints on the triplet VEV than in the GM model and enhancing the single production cross section at hadron and lepton colliders.

    \item The region with $m_{F^{++}}<200$ GeV is not directly constrained by dedicated LHC searches~\cite{Ashanujjaman:2022ofg}. Our global fit places indirect limits through searches for other BSM scalars, highlighting this region as an important focus area for future colliders. In particular, the eGM model offers additional search channels through the above-mentioned cascade decays.

\end{itemize}
Thus, the relatively light mass spectra and new decay modes provide well-motivated benchmarks for future collider studies. Dedicated searches for these collider signatures, together with precision measurements of 125 GeV Higgs boson couplings, will play a crucial role in testing the GM and eGM models at the HL-LHC and future colliders. This motivates a detailed collider study and will be carried out in future work.

\begin{acknowledgments}
We thank Debtosh Chowdhury, Anirban Kundu, and 	Saiyad Ashanujjaman for useful discussions and their valuable comments on the manuscript. P.M. acknowledges financial support from the ANRF, Government of India, under Grant No.~CRG/2021/007579. S.S. acknowledges financial support from the MHRD, Government of India, under the PMRF Scheme.
\end{acknowledgments}

\appendix
\section{}\label{app:A}
In this appendix, we list the experimental inputs used in our global fit. These are the signal strengths of the 125~GeV $h$ (Tables~\ref{tab:signalstrengthsAC78}--\ref{tab:signalstrengthsA13}), the direct searches for light neutral scalars (Tables~\ref{tab:ATLAS_light} and~\ref{tab:CMS_light}), the direct searches for heavy neutral scalars (Tables~\ref{tab:CMS_heavy_scalar_to_scalar}--\ref{tab:ATLAS_heavy_scalar_to_fermions}), and the searches for light charged scalars in top-quark decays (Tables~\ref{tab:CMS_charged} and~\ref{tab:ATLAS_charged}).

\begin{table}[ht]
\begin{center}
\setlength{\tabcolsep}{5pt}
\renewcommand{\arraystretch}{1.3}
\scalebox{0.85}{
\begin{tabular}{ | c c  cccccc  |}
\hline
Signal Strength & Value & \multicolumn{6}{c|}{Correlation Matrix} \\
\hline
$\mu^{\gamma\gamma}_{ggF}$ & $1.10 \pm 0.23$ & $1$     & $-0.25$ & $0$     & $-0.14$ & $0$     & \\
$\mu^{\gamma\gamma}_{VBF}$ & $1.3 \pm 0.5$   & $-0.25$ & $1$     & $0$     & $0$     & $0$     & \\
$\mu^{\gamma\gamma}_{Wh}$  & $0.5 \pm 1.3$   & $0$     & $0$     & $1$     & $-0.64$ & $0$     & \\
$\mu^{\gamma\gamma}_{Zh}$  & $0.5 \pm 2.8$   & $-0.14$ & $0$     & $-0.64$ & $1$     & $-0.11$ & \\
$\mu^{\gamma\gamma}_{tth}$ & $2.2 \pm 1.5$   & $0$     & $0$     & $0$     & $-0.11$ & $1$     & \\
\hline
$\mu^{ZZ}_{ggF}$ & $1.13 \pm 0.33$ & $1$     & $-0.26$ & & & & \\
$\mu^{ZZ}_{VBF}$ & $0.1 \pm 0.9$   & $-0.26$ & $1$     & & & & \\
\hline
$\mu^{WW}_{ggF}$ & $0.84 \pm 0.17$ & $1$     & $-0.16$ & & & & \\
$\mu^{WW}_{VBF}$ & $1.2 \pm 0.4$   & $-0.16$ & $1$     & & & & \\
\hline
$\mu^{\tau\tau}_{ggF}$ & $1.0 \pm 0.6$  & $1$     & $-0.37$ & $0$     & $-0.25$ & $0$     & $-0.21$ \\
$\mu^{\tau\tau}_{VBF}$ & $1.3 \pm 0.4$  & $-0.37$ & $1$     & $0$     & $0$     & $0$     & $0$     \\
$\mu^{WW}_{Wh}$        & $1.6 \pm 1.1$  & $0$     & $0$     & $1$     & $-0.12$ & $-0.12$ & $0$     \\
$\mu^{\tau\tau}_{Wh}$  & $-1.4 \pm 1.4$ & $-0.25$ & $0$     & $-0.12$ & $1$     & $0$     & $0$     \\
$\mu^{WW}_{Zh}$        & $5.9 \pm 2.4$  & $0$     & $0$     & $-0.12$ & $0$     & $1$     & $0$     \\
$\mu^{\tau\tau}_{Zh}$  & $2.2 \pm 2.0$  & $-0.21$ & $0$     & $0$     & $0$     & $0$     & $1$     \\
\hline
$\mu^{WW}_{tth}$       & $5.0 \pm 1.8$  & $1$     & $-0.47$ & & & & \\
$\mu^{\tau\tau}_{tth}$ & $-1.9 \pm 3.5$ & $-0.47$ & $1$     & & & & \\
\hline
$\mu^{bb}_{Wh}$    & $1.0 \pm 0.5$ & & & & & & \\
$\mu^{bb}_{Zh}$    & $0.4 \pm 0.4$ & & & & & & \\
$\mu^{bb}_{tth}$   & $1.1 \pm 1.0$ & & & & & & \\
$\mu^{\mu\mu}_{pp}$ & $0.1 \pm 2.5$ & & & & & & \\
\hline
\end{tabular}}
\caption{Summary of the $h$ signal strengths and their correlations from the official ATLAS and CMS combination of the Run~1 dataset at $\sqrt{s} = 7$ and $8$~TeV~\cite{Chowdhury:2017aav}.}
\label{tab:signalstrengthsAC78}
\end{center}
\end{table}
\begin{table}[ht]
\begin{center}
\setlength{\tabcolsep}{5pt}
\renewcommand{\arraystretch}{1.3}
\scalebox{0.85}{
\begin{tabular}{ | c c  cccc  |}
\hline
 \;Signal Strength & \;Value\; & \multicolumn{4}{c|}{\;Correlation Matrix\;\;} \\
\hline
  $\mu_\text{ggh,bbh}^{\gamma \gamma}$ &\; $1.07 \pm 0.11$ \; &&&&  \\[2pt]
  $\mu_\text{VBF}^{\gamma \gamma}$ & $1.04 \pm 0.32$  &&&& \\[2pt]
  $\mu_\text{Vh}^{\gamma \gamma}$ & $1.34 \pm 0.34$  &&&& \\[2pt]
  $\mu_\text{tth,th}^{\gamma \gamma}$ & $1.35 \pm 0.31$  &&&& \\[3pt]
\hline
  $\;\mu_\text{ggh,bbh,tth,th}^{ZZ}$\; & $0.95 \pm 0.13$  &1&$-0.11$&&\\[2pt]
  $\mu_\text{VBF,Vh}^{ZZ}$ & $0.82 \pm 0.34$  &$-0.11$&1&& \\[3pt]
\hline
  $\mu_\text{ggh}^{WW}$ & $0.92 \pm 0.11$  &1 &\;$-0.13$ \;\;&0 & 0  \\[2pt]
  $\mu_\text{VBF}^{WW}$ & $0.71 \pm 0.26$  &\;$-0.13$ &1 & 0&0   \\[2pt]
  $\mu_\text{Zh}^{WW}$ & $2.0\pm 0.7$  &0&0&1&0  \\[2pt]
  $\mu_\text{Wh}^{WW}$ & $2.2\pm 0.6$  &0&0&0&1  \\[3pt]
\hline
  $\mu_\text{incl.}^{\tau \tau}$ & $0.93 \pm 0.12$  &&&&  \\[2pt]
  $\mu_\text{ggh}^{\tau \tau}$ & $0.97 \pm 0.19$  &&&&   \\[2pt]
  $\mu_\text{qqh}^{\tau \tau}$ & $0.68 \pm 0.23$  &&&& \\[2pt]
  $\mu_\text{Vh}^{\tau \tau}$ & $1.80 \pm 0.44$  &&&&  \\[3pt]
\hline
  $\mu_\text{qqh}^{bb}$ & $1.59 \pm 0.60$  &1 &$-0.75$ & & \\[2pt]
  $\mu_\text{ggh}^{bb}$ & $-2.7 \pm 3.89$  &$-0.75$ &1 & &  \\[3pt]
\hline
  $\mu_\text{ggh,tth}^{\mu\mu}$ & $0.66 \pm 0.67$  & 1&$-0.24$ & &  \\[2pt]
  $\mu_\text{VBF,Vh}^{\mu\mu}$ & $1.85 \pm 0.86$  &$-0.24$ &1 & & \\[3pt]
\hline
  $\mu_\text{pp}^{Z\gamma}$ & $2.4 \pm 0.9$  & & & & \\[3pt]
\hline
\end{tabular}}
\caption{Summary of the $h$ signal strengths obtained by CMS from the full Run 2 dataset at 13 TeV~\cite{Chowdhury:2024mfu}.}
\label{tab:signalstrengthsC13}
\end{center}
\end{table}
\begin{table}[h!]
\begin{center}
\setlength{\tabcolsep}{5pt}
\renewcommand{\arraystretch}{1.25}
\scalebox{0.8}{
\begin{tabular}{ | c  c  cccccc |}

\hline
 \;Signal Strength & \;Value\; & \multicolumn{6}{c|}{\;Correlation Matrix\;\;} \\
\hline
 $\mu_\text{ggF,bbh}^{\gamma \gamma}$ &$1.04\pm 0.10$  &\;1\;&$\;-0.13\;\;$&0&0&0&0 \\[2pt]
 $\mu_\text{VBF}^{\gamma \gamma}$ & $1.20 \pm 0.26$  &\;$-0.13$\;&1&0&0&0&0 \\[2pt]
  $\mu_\text{Wh}^{\gamma \gamma}$ & $1.5 \pm 0.55$  &0&0&1&\;$-0.37$\;\;&0&\;$-0.11$\;\; \\[2pt]
  $\mu_\text{Zh}^{\gamma \gamma}$ & $-0.2 \pm 0.55$  &0&0&\;$-0.37$\;\;&1&0&0 \\[2pt]
  $\mu_\text{tth}^{\gamma \gamma}$ & $0.89 \pm 0.31$  &0&0&0&0&1&$-0.44$ \\[2pt]
  $\mu_\text{th}^{\gamma \gamma}$ & $3 \pm 3.5$  &0&0&$-0.11$&0&$-0.44$&1 \\[3pt]
\hline
  $\mu_\text{ggF}^{ZZ}$ & $0.95 \pm 0.1$  &1&$-0.22$&$-0.27$&0& &  \\[2pt]
  $\mu_\text{VBF}^{ZZ}$ & $1.19 \pm 0.45 $  &$-0.22$&1&0&0&& \\[2pt]
 $\mu_\text{Vh}^{ZZ}$ & $1.43 \pm 1.0$  &$-0.27$&0&1&$-0.18$ && \\[2pt]
  $\mu_\text{tth}^{ZZ}$ & $1.69 \pm 1.45$  &0&0&$-0.18$&1 && \\[3pt]
\hline
  $\mu_\text{incl.}^{ZZ}$ & $1.0 \pm 0.1$  &&&& &&  \\[3pt]
\hline
  $\mu_\text{ggF,bbh}^{WW}$ & $1.15 \pm 0.135$  & & & & & & \\[2pt]
  $\mu_\text{VBF}^{WW}$ & $0.93 \pm 0.21$  & & & & & &  \\[2pt]
  $\mu_\text{ggF,bbh,VBF}^{WW}$ & $1.09 \pm 0.11$  &&&&&&  \\[3pt]
\hline
  $\mu_\text{VBF}^{\tau \tau}$ & $0.90 \pm 0.18$  &1&$-0.24$&0&0&&  \\[2pt]
  $\mu_\text{ggF,bbh}^{\tau \tau}$ & $0.96 \pm 0.31$  &$-0.24$&1&$-0.29$&0&&   \\[2pt]
  $\mu_\text{Vh}^{\tau \tau}$ & $0.98 \pm 0.60$  &0&$-0.29$&1&0&& \\[2pt]
  $\mu_\text{tth,th}^{\tau \tau}$ & $1.06 \pm 1.18$  &0&0&0&1&&  \\[3pt]
\hline
  $\mu_\text{VBF}^{bb}$ & $0.95 \pm 0.37$  & & & & & & \\[2pt]
  $\mu_\text{Wh}^{bb}$ & $0.95 \pm 0.26$  & & & & & &  \\[2pt]
  $\mu_\text{Zh}^{bb}$ & $1.08 \pm 0.24$  & & & & & &  \\[2pt]
  $\mu_\text{Vh}^{bb}$ & $1.02 \pm 0.17$  & & & & & &  \\[2pt]
  $\mu_\text{tth,th}^{bb}$ & $0.35 \pm 0.35$  & & & & & &  \\[3pt]
\hline
  $\mu_\text{pp}^{\mu\mu}$ & $1.2\pm 0.6$  & & & & & &  \\[3pt]
\hline
  $\mu_\text{pp}^{Z\gamma}$ & $2.0 \pm 0.95$  & & & & & &  \\[3pt]
\hline
\end{tabular}}
\caption{Summary of the $h$ signal strengths obtained by ATLAS from the full Run 2 dataset at 13 TeV~\cite{Chowdhury:2024mfu}.}
\label{tab:signalstrengthsA13}
\end{center}
\end{table}

\begin{table}[ht!]
\centering
\small
\renewcommand{\arraystretch}{1.2}
\begin{tabular}{l c c}
\hline\hline
Channel & Mass range (GeV) & Reference \\
\hline
$pp\to h \to AA \to \tau^+\tau^-\tau^+\tau^-$ & $[3.7,50]$ & \cite{ATLAS:2015unc} \\
$pp\to h \to AA \to \gamma\gamma\gamma\gamma$ & $[10,62]$ & \cite{ATLAS:2015rsn} \\
\hline
$pp\to h \to \varphi_i^0\varphi_i^0 \to \gamma\gamma\, gg$ & $[20,60]$ & \cite{ATLAS:2018jnf} \\
$pp\to h(V) \to \varphi_i^0\varphi_i^0(\ell^+\ell^-) \to b\bar{b}\,b\bar{b}\,(\ell^+\ell^-)$ & $[20,60]$ & \cite{ATLAS:2018pvw} \\
$pp\to h \to \varphi_i^0\varphi_i^0 \to \mu^+\mu^-\, b\bar{b}$ & $[20,60]$ & \cite{ATLAS:2018emt} \\
$pp\to h(Z) \to \varphi_i^0\varphi_i^0(\ell^+\ell^-) \to b\bar{b}\,b\bar{b}\,(\ell^+\ell^-)$ & $[15,30]$ & \cite{ATLAS:2020ahi} \\
$pp\to h \to AA \to \mu^+\mu^-\, b\bar{b}$ & $[16,62]$ & \cite{ATLAS:2021hbr} \\
\hline
$pp\to h \to \varphi_i^0(Z) \to \mu^+\mu^-(\ell^+\ell^-)$ & $[15,30]$ & \multirow{2}{*}{\cite{ATLAS:2021ldb}} \\
$pp\to h \to \varphi_i^0\varphi_i^0 \to \mu^+\mu^-\mu^+\mu^-$ & $[1,60]$ & \\
\hline
$pp\to A(t\bar{t}) \to \mu^+\mu^-(t\bar{t})$ & $[15,72]$ & \cite{ATLAS:2023ofo} \\
$pp\to h \to AA \to b\bar{b}\,\tau^+\tau^-$ & $[12,60]$ & \cite{ATLAS:2024vpj} \\
$pp\to h \to AA \to \gamma\gamma\gamma\gamma$ & $[10,62]$ & \cite{ATLAS:2023ian} \\
$gg\to A \to \tau^+\tau^-$ & $[20,90]$ & \cite{ATLAS:2024rzd} \\
$pp\to H \to \gamma\gamma$ & $[66,110]$ & \cite{ATLAS:2024bjr} \\
\hline\hline
\end{tabular}
\caption{ATLAS searches for neutral eGM/GM scalars, $\varphi_i^0 = A/ H_3^0$ (taken from Ref.~\cite{Coutinho:2024zyp}).}
\label{tab:ATLAS_light}
\end{table}

\begin{table}[ht!]
\centering
\small
\renewcommand{\arraystretch}{1.2}
\begin{tabular}{l c c}
\hline\hline
Channel & Mass range (GeV) & Reference \\
\hline
$pp\to A(b\bar{b}) \to \tau^+\tau^-(b\bar{b})$ & $[25,80]$ & \cite{CMS:2015qnd} \\
$pp\to h \to AA \to \tau^+\tau^-\tau^+\tau^-$ & $[5,15]$ & \multirow{3}{*}{\cite{CMS:2017dmg}} \\
\hline
$pp\to h \to AA \to \mu^+\mu^-\, b\bar{b}$ & $[25,62.5]$ & \\
$pp\to h \to AA \to \mu^+\mu^-\tau^+\tau^-$ & $[15,62.5]$ & \\
\hline
$pp\to A(b\bar{b}) \to \mu^+\mu^-(b\bar{b})$ & $[25,60]$ & \cite{CMS:2017nmj} \\
$pp\to H \to \gamma\gamma$ & $[80,110]$ & \cite{CMS:2018cyk} \\
\hline
$pp\to h \to AA \to \mu^+\mu^-\tau^+\tau^-$ & $[15,62.5]$ & \cite{CMS:2018qvj} \\
$pp\to h \to AA \to \tau^+\tau^-\, b\bar{b}$ & $[12,60]$ & \cite{CMS:2024uru} \\
$pp\to h \to AA \to \mu^+\mu^-\, b\bar{b}$ & $[15,62]$ & \cite{CMS:2024uru} \\
$pp\to A(b\bar{b}) \to \tau^+\tau^-(b\bar{b})$ & $[25,70]$ & \cite{CMS:2019hvr} \\
$pp\to h \to AA \to \tau^+\tau^-\tau^+\tau^-$ & $[4,15]$ & \cite{CMS:2019spf} \\
\hline
$pp\to h \to \varphi_i^0(Z) \to \ell^+\ell^-(\ell^+\ell^-)$ & \multirow{2}{*}{$[4,60]$} & \multirow{2}{*}{\cite{CMS:2021pcy}} \\
$pp\to h \to \varphi_i^0\varphi_i^0 \to \ell^+\ell^-\ell^+\ell^-$ & & \\
\hline
$pp\to h \to AZ \to \gamma\gamma\,\ell^+\ell^-$ & $[1,30]$ & \cite{CMS:2023eos} \\
$pp\to h \to AA \to \gamma\gamma\gamma\gamma$ & $[15,62]$ & \cite{CMS:2022xxa} \\
$pp\to h \to AA \to b\bar{b}\,b\bar{b}$ & $[15,60]$ & \cite{CMS:2024zfv} \\
\hline
$pp\to \varphi_i^0(V) \to \ell^+\ell^-(\ell\nu/\ell^+\ell^-)$ & \multirow{2}{*}{$[15,350]$} & \multirow{2}{*}{\cite{CMS:2024ulc}} \\
$pp\to \varphi_i^0(t\bar{t}) \to \ell^+\ell^-(t\bar{t})$ & & \\
\hline
$pp\to H \to \gamma\gamma$ & $[70,110]$ & \cite{CMS:2024yhz} \\
\hline\hline
\end{tabular}
\caption{CMS searches for neutral eGM/GM scalars, $\varphi_i^0 = A/ H_3^0$ (taken from Ref.~\cite{Coutinho:2024zyp}).}
\label{tab:CMS_light}
\end{table}

\begin{table}[ht!]
\centering
\small
\renewcommand{\arraystretch}{1.2}
\begin{tabular}{l c c}
\hline\hline
Channel & Mass range (TeV) & Reference \\
\hline
$pp\to \varphi_i^0 \to hh \to (bb)(bb)$ & $[0.27,1.1]$ & \cite{CMS:2015jal} \\
$pp\to \varphi_i^0 \to hh \to (bb)(\gamma\gamma)$ & $[0.26,1.1]$ & \cite{CMS:2016cma} \\
$gg\to \varphi_i^0 \to hh \to (bb)(\tau\tau)$ & $[0.26,0.35]$ & \cite{CMS:2015uzk} \\
$pp\to \varphi_i^0 \to hh [\to (bb)(\tau\tau)]$ & $[0.35,1]$ & \cite{CMS:2017yfv} \\
$gg\to \varphi_i^0 \to hZ \to (bb)(\ell\ell)$ & $[0.225,0.6]$ & \cite{CMS:2015flt} \\
$gg\to \varphi_i^0 \to hZ \to (\tau\tau)(\ell\ell)$ & $[0.22,0.35]$ & \cite{CMS:2015uzk} \\
\hline
$pp\to \varphi_3^0 \to \varphi_2^0 Z \to (bb)(\ell\ell)$ & $0.04 < M_{\varphi_2^0} < 1.0$ & \cite{CMS:2016xnc} \\
$pp\to \varphi_3^0 \to \varphi_2^0 Z \to (\tau\tau)(\ell\ell)$ & $0.05 < M_{\varphi_2^0} < 1.0$ & \cite{CMS:2016xnc} \\
\hline
\multirow{2}{*}{$pp\to \varphi_i^0 \to hh \to (bb)(bb)$} & $[0.26,1.2]$ & \cite{CMS:2018qmt} \\
 & $[1,3]$ & \cite{CMS:2024pjq} \\
 \hline
$pp\to \varphi_i^0 \to hh [\to (WW)(WW)/(WW)(\tau\tau)/(\tau\tau)(\tau\tau)]$ & $[0.25,1]$ & \cite{CMS:2022kdx} \\
$pp\to \varphi_i^0 \to hh \to (bb)(\gamma\gamma)$ & $[0.26,1]$ & \cite{CMS:2023boe} \\
$pp\to \varphi_i^0 \to hh \to (bb)(\tau\tau)$ & $[0.25,0.9]$ & \cite{CMS:2017hea} \\
$pp\to \varphi_i^0 \to hh [\to (bb)(\tau\tau)]$ & $[0.9,4]$ & \cite{CMS:2018kaz} \\
$pp\to \varphi_i^0 \to hh \to (bb)(VV\to \ell\nu\ell\nu)$ & $[0.26,0.9]$ & \cite{CMS:2017rpp} \\
$pp\to \varphi_i^0 \to hh [\to (bb)(WW\to q\bar{q}\ell\nu)]$ & $[0.8,3.5]$ & \cite{CMS:2019noi} \\
$pp\to \varphi_i^0 \to hh \to (bb)[ZZ\to \ell\ell jj]$ & $[0.26,1]$ & \cite{CMS:2020jeo} \\
$pp\to \varphi_i^0 \to hh \to (bb)[ZZ\to \ell\ell\nu\nu]$ & $[0.26,1]$ & \cite{CMS:2020jeo} \\
\hline
\multirow{2}{*}{$pp\to \varphi_i^0 \to hh [\to (bb)(WW/\tau\tau\to (q\bar{q}/\ell\nu)\ell\nu)]$} & $[0.8,4.5]$ & \cite{CMS:2021roc} \\
 & $[0.25,0.9]$ & \cite{CMS:2024rgy} \\
 \hline
\multirow{2}{*}{$gg\to \varphi_i^0 \to (h\to b\bar{b})[Z\to \nu\bar{\nu}/\ell\bar{\ell}]$} & $[0.22,0.8]$ & \cite{CMS:2019qcx} \\
 & $[0.8,2]$ & \cite{CMS:2018ljc} \\
 \hline
$gg\to \varphi_i^0 \to (h\to\tau\tau)(Z\to\ell\ell)$ & $[0.22,0.4]$ & \cite{CMS:2019kca} \\
\hline
\multirow{2}{*}{$bb\to \varphi_i^0 \to (h\to b\bar{b})[Z\to \nu\bar{\nu}/\ell\bar{\ell}]$} & $[0.22,0.8]$ & \cite{CMS:2019qcx} \\
 & $[0.8,2]$ & \cite{CMS:2018ljc} \\
\hline\hline
\end{tabular}
\caption{CMS searches for the neutral eGM/GM scalars, $\varphi_i^0 = A,F^0/ H_3^0, H_5^0$, decaying into $h$ or a lighter neutral scalar (taken from Ref.~\cite{Coutinho:2024zyp}). Here, $\varphi_3^0$: heaviest neutral scalar; $V \in \{W, Z\}$; $\ell \in \{e, \mu\}$.}
\label{tab:CMS_heavy_scalar_to_scalar}
\end{table}


\begin{table}[ht!]
\centering
\small
\renewcommand{\arraystretch}{1.2}
\begin{tabular}{l c c}
\hline\hline
Channel & Mass range (TeV) & Reference \\
\hline
$gg\to \varphi_i^0 \to hh$ & $[0.26,1]$ & \cite{ATLAS:2015sxd} \\
$gg\to \varphi_i^0 \to hZ \to (bb)Z$ & $[0.22,1]$ & \cite{ATLAS:2015kpj} \\
$gg\to \varphi_i^0 \to hZ \to (\tau\tau)Z$ & $[0.05,1]$ & \cite{ATLAS:2015kpj} \\
\hline
$pp\to \varphi_i^0 \to hh [\to (bb)(bb)]$ & $[0.251,5]$ & \cite{ATLAS:2022hwc} \\
$pp\to \varphi_i^0 \to hh [\to (bb)(\gamma\gamma)]$ & $[0.251,1]$ & \cite{ATLAS:2021ifb} \\
\hline
\multirow{2}{*}{$pp\to \varphi_i^0 \to hh [\to (bb)(\tau\tau)]$} & $[0.251,1.6]$ & \cite{ATLAS:2022xzm} \\
 & $[1,3]$ & \cite{ATLAS:2020azv} \\
 \hline
$pp\to \varphi_i^0 \to hh [\to (bb)(WW)]$ & $[0.5,3]$ & \cite{ATLAS:2018fpd} \\
$gg\to \varphi_i^0 \to hh \to (\gamma\gamma)(WW)$ & $[0.26,0.5]$ & \cite{ATLAS:2018hqk} \\
\hline
$gg\to \varphi_i^0 \to hZ [\to (bb)Z]$ & \multirow{2}{*}{$[0.22,2]$} & \multirow{2}{*}{\cite{ATLAS:2022enb}} \\
$bb\to \varphi_i^0 \to hZ [\to (bb)Z]$ & & \\
\hline
$gg\to \varphi_3^0 \to \varphi_2^0 Z \to (bb)Z$ & $0.13 < M_{\varphi_3^0} < 0.7$ & \multirow{3}{*}{\cite{ATLAS:2020gxx}} \\
$bb\to \varphi_3^0 \to \varphi_2^0 Z \to (bb)Z$ & $0.23 < M_{\varphi_2^0} < 0.8$ & \\
\hline
$gg\to \varphi_3^0 \to \varphi_2^0 Z \to (WW)Z$ & $0.2 < M_{\varphi_3^0} < 0.7$ / $0.3 < M_{\varphi_2^0} < 0.8$ &\cite{ATLAS:2020gxx} \\
\hline\hline
\end{tabular}
\caption{ATLAS searches for $\varphi_i^0$ decaying into $h$ or a lighter neutral scalar (taken from Ref.~\cite{Coutinho:2024zyp}). The notation follows Table~\ref{tab:CMS_heavy_scalar_to_scalar}. }
\label{tab:ATLAS_heavy_scalar_to_scalar}
\end{table}


\begin{table}[ht!]
\centering
\small
\renewcommand{\arraystretch}{1.2}
\begin{tabular}{l c c}
\hline\hline
Channel & Mass range (TeV) & Reference \\
\hline
$pp\to \varphi_i^0 \to Z\gamma \to (\ell\ell)\gamma$ & $[0.2,1.2]$ & \cite{CMS:2016all} \\
$pp\to \varphi_i^0 \to VV$ & $[0.145,1]$ & \cite{CMS:2015hra} \\
\hline
$gg\to \varphi_i^0 \to \gamma\gamma$ & $[0.6,5]$ & \cite{CMS:2024nht} \\
$pp\to \varphi_i^0 \to Z\gamma [\to (\ell\ell \,\&\, qq)\gamma]$ & $[0.35,4]$ & \cite{CMS:2017dyb} \\
$pp\to \varphi_i^0 \to ZZ [\to (\ell\ell)(qq,\nu\nu,\ell\ell)]$ & $[0.13,3]$ & \cite{CMS:2018amk} \\
$pp\to \varphi_i^0 \to ZZ [\to (qq)(\nu\nu)]$ & $[1,4]$ & \cite{CMS:2018ygj} \\
\hline
$VV\to \varphi_i^0 \to WW$ & \multirow{2}{*}{$[0.2,3]$} & \multirow{2}{*}{\cite{CMS:2019bnu}} \\
$gg\to \varphi_i^0 \to WW$ & & \\
\hline
$pp\to \varphi_i^0 \to WW [\to (\ell\nu)(qq)]$ & $[1,4.4]$ & \cite{CMS:2018dff} \\
\hline
$gg\to \varphi_i^0 \to WW$ & \multirow{2}{*}{$[1,4.5]$} & \multirow{2}{*}{\cite{CMS:2021klu}} \\
$VV\to \varphi_i^0 \to WW$ & & \\
\hline
$(gg\!+\!VV)\to \varphi_i^0 \to WW \to (\ell\nu)(\ell\nu)$ & $[0.2,1]$ & \cite{CMS:2016jpd} \\
\hline\hline
\end{tabular}
\caption{CMS searches for $\varphi_i^0$ decaying into gauge bosons (taken from Ref.~\cite{Coutinho:2024zyp}). The notation follows Table~\ref{tab:CMS_heavy_scalar_to_scalar}. }
\label{tab:CMS_heavy_scalar_to_bosons}
\end{table}


\begin{table}[ht!]
\centering
\small
\renewcommand{\arraystretch}{1.2}
\begin{tabular}{l c c}
\hline\hline
Channel & Mass range (TeV) & Reference \\
\hline
$gg\to \varphi_i^0 \to \gamma\gamma$ & $[0.065,0.6]$ & \cite{ATLAS:2014jdv} \\
$pp\to \varphi_i^0 \to Z\gamma \to (\ell\ell)\gamma$ & $[0.2,1.6]$ & \cite{ATLAS:2014lfk} \\
$gg\to \varphi_i^0 \to ZZ$ & $[0.14,1]$ & \cite{ATLAS:2015pre} \\
\hline
$gg\to \varphi_i^0 \to WW$ & \multirow{2}{*}{$[0.3,1.5]$} & \multirow{2}{*}{\cite{ATLAS:2015iie}} \\
$VV\to \varphi_i^0 \to WW$ & & \\
\hline
$pp\to \varphi_i^0 \to \gamma\gamma$ & $[0.15,3]$ & \cite{ATLAS:2021uiz} \\
$gg\to \varphi_i^0 \to Z\gamma [\to (\ell\ell)\gamma]$ & $[0.22,3.4]$ & \cite{ATLAS:2023wqy} \\
$gg\to \varphi_i^0 \to Z\gamma [\to (qq)\gamma]$ & $[1.02,6.8]$ & \cite{ATLAS:2023kcu} \\
\hline
$gg\to \varphi_i^0 \to ZZ [\to (\ell\ell)(\ell\ell,\nu\nu)]$ & \multirow{2}{*}{$[0.2,2]$} & \multirow{2}{*}{\cite{ATLAS:2020tlo}} \\
$VV\to \varphi_i^0 \to ZZ [\to (\ell\ell)(\ell\ell,\nu\nu)]$ & & \\
\hline
$gg\to \varphi_i^0 \to ZZ [\to (\ell\ell,\nu\nu)(qq)]$ & \multirow{2}{*}{$[0.3,3]$} & \multirow{2}{*}{\cite{ATLAS:2017otj}} \\
$VV\to \varphi_i^0 \to ZZ [\to (\ell\ell,\nu\nu)(qq)]$ & & \\
\hline
$gg\to \varphi_i^0 \to WW [\to (e\nu)(\mu\nu)]$ & \multirow{2}{*}{$[0.2,4]$} & \multirow{2}{*}{\cite{ATLAS:2022qlc}} \\
$VV\to \varphi_i^0 \to WW [\to (e\nu)(\mu\nu)]$ & & \\
\hline
$gg\to \varphi_i^0 \to WW [\to (\ell\nu)(qq)]$ & \multirow{2}{*}{$[0.3,3]$} & \multirow{2}{*}{\cite{ATLAS:2017jag}} \\
$VV\to \varphi_i^0 \to WW [\to (\ell\nu)(qq)]$ & & \\
\hline
$pp\to \varphi_i^0 \to VV [\to (qq)(qq)]$ & $[1.2,3]$ & \cite{ATLAS:2017zuf} \\
\hline
$gg\to \varphi_i^0 \to VV$ & \multirow{2}{*}{$[0.2,5.2]$} & \multirow{2}{*}{\cite{ATLAS:2020fry}} \\
$VV\to \varphi_i^0 \to VV$ & & \\
\hline\hline
\end{tabular}
\caption{ATLAS searches for $\varphi_i^0$ decaying into gauge bosons (taken from Ref.~\cite{Coutinho:2024zyp}). The notation follows Table~\ref{tab:CMS_heavy_scalar_to_scalar}.}
\label{tab:ATLAS_heavy_scalar_to_bosons}
\end{table}


\begin{table}[ht!]
\centering
\small
\renewcommand{\arraystretch}{1.2}
\begin{tabular}{l c c}
\hline\hline
Channel & Mass range (TeV) & Reference \\
\hline
$bb\to \varphi_i^0 \to bb$ & $[0.1,0.9]$ & \cite{CMS:2015grx} \\
$gg\to \varphi_i^0 \to bb$ & $[0.33,1.2]$ & \cite{CMS:2018kcg} \\
\hline
$bb\to \varphi_i^0 \to \mu\mu$ & \multirow{2}{*}{$[0.12,0.5]$} & \multirow{2}{*}{\cite{CMS:2015ooa}} \\
$gg\to \varphi_i^0 \to \mu\mu$ & & \\
\hline
$bb\to \varphi_i^0 \to \tau\tau$ & \multirow{2}{*}{$[0.09,1]$} & \multirow{2}{*}{\cite{CMS:2015mca}} \\
$gg\to \varphi_i^0 \to \tau\tau$ & & \\
\hline
$tt/tW/tq\to \varphi_i^0 \to tt$ & $[0.35,0.65]$ & \cite{CMS:2019rvj} \\
\hline
\multirow{2}{*}{$pp\to \varphi_i^0 \to bb$} & $[0.55,1.2]$ & \cite{CMS:2016ncz} \\
 & $[0.05,0.35]$ & \cite{CMS:2018pwl} \\
 \hline
$bb\to \varphi_i^0 \to bb$ & $[0.3,1.3]$ & \cite{CMS:2018hir} \\
\hline
$bb\to \varphi_i^0 \to \mu\mu$ & \multirow{2}{*}{$[0.14,1]$} & \multirow{2}{*}{\cite{CMS:2019mij}} \\
$gg\to \varphi_i^0 \to \mu\mu$ & & \\
\hline
$bb\to \varphi_i^0 \to \tau\tau$ & \multirow{2}{*}{$[0.06,3.5]$} & \multirow{2}{*}{\cite{CMS:2022goy}} \\
$gg\to \varphi_i^0 \to \tau\tau$ & & \\
\hline\hline
\end{tabular}
\caption{CMS searches for $\varphi_i^0=A,F^0/ H_3^0, H_5^0$ decaying into quarks or leptons (taken from Ref.~\cite{Coutinho:2024zyp}).}
\label{tab:CMS_heavy_scalar_to_fermions}
\end{table}


\begin{table}[ht!]
\centering
\small
\renewcommand{\arraystretch}{1.2}
\begin{tabular}{l c c}
\hline\hline
Channel & Mass range (TeV) & Reference \\
\hline
$gg\to \varphi_i^0 \to \tau\tau$ & \multirow{2}{*}{$[0.09,1]$} & \multirow{2}{*}{\cite{ATLAS:2014vhc}} \\
$bb\to \varphi_i^0 \to \tau\tau$ & & \\
\hline
$bb\to \varphi_i^0 \to tt$ & \multirow{2}{*}{$[0.4,1]$} & \cite{ATLAS:2018cye} \\
$tt\to \varphi_i^0 \to tt$ & & \cite{ATLAS:2022rws} \\
\hline
$pp\to \varphi_i^0 \to bb$ ($\geq 1$ $b$-jet) & $[1.4,6.6]$ & \multirow{3}{*}{\cite{ATLAS:2018tfk}} \\
\hline
\multirow{2}{*}{$pp\to \varphi_i^0 \to bb$} & $[0.6,1.25]$ & \\
 & $[1.25,6.2]$ & \\
 \hline
$bb\to \varphi_i^0 \to bb$ & $[0.45,1.4]$ & \cite{ATLAS:2019tpq} \\
\hline
$bb\to \varphi_i^0 \to \mu\mu$ & \multirow{2}{*}{$[0.2,1]$} & \multirow{2}{*}{\cite{ATLAS:2019odt}} \\
$gg\to \varphi_i^0 \to \mu\mu$ & & \\
\hline
$bb\to \varphi_i^0 \to \tau\tau$ & \multirow{2}{*}{$[0.2,2.5]$} & \multirow{2}{*}{\cite{ATLAS:2020zms}} \\
$gg\to \varphi_i^0 \to \tau\tau$ & & \\
\hline\hline
\end{tabular}
\caption{ATLAS searches for $\varphi_i^0=A,F^0/ H_3^0, H_5^0$ decaying into quarks or leptons (taken from Ref.~\cite{Coutinho:2024zyp}).}
\label{tab:ATLAS_heavy_scalar_to_fermions}
\end{table}
\begin{table}[ht!]
\centering
\small
\renewcommand{\arraystretch}{1.2}
\begin{tabular}{l c c}
\hline\hline
Channel & Mass range (GeV) & Reference \\
\hline
$t\to (H^+\to \tau^+\nu)\, b$ & $[80,160]$ & \cite{CMS:2015lsf} \\
$t\to (H^+\to c\bar{s} = 100\%)\, b$ & $[90,160]$ & \cite{CMS:2015yvc} \\
$t\to (H^+\to c\bar{b} = 100\%)\, b$ & $[90,150]$ & \cite{CMS:2018dzl} \\
\hline
$t\to b\,(H^+\to W^+A)\to b\,(W^+\mu^+\mu^-)$ & $\begin{matrix} 15 < M_A < 75 \\ 100 < M_{H^+} < 160 \end{matrix}$ & \cite{CMS:2019idx} \\
\hline
$t\to (H^+\to c\bar{s} = 100\%)\, b$ & $[80,160]$ & \cite{CMS:2020osd} \\
\hline\hline
\end{tabular}
\caption{CMS searches for light charged scalars (taken from Ref.~\cite{Coutinho:2024zyp}).}
\label{tab:CMS_charged}
\end{table}


\begin{table}[ht!]
\centering
\small
\renewcommand{\arraystretch}{1.2}
\begin{tabular}{l c c}
\hline\hline
Channel & Mass range (GeV) & Reference \\
\hline
$t\to (H^+\to \tau^+\nu)\, b$ & $[80,160]$ & \cite{ATLAS:2014otc} \\
\hline
$t\to (H^+\to c\bar{b})\, b$ & $[60,160]$ & \cite{ATLAS:2023bzb} \\
\hline
$t\to b\,(H^+\to W^+A)\to b\,(W^+\mu^+\mu^-)$ & $\begin{matrix} 15 < M_A < 72 \\ 120 < M_{H^+} < 160 \end{matrix}$ & \cite{ATLAS:2023ofo} \\
\hline
$t\to (H^+\to c\bar{s})\, b$ & $[60,168]$ & \cite{ATLAS:2024oqu} \\
\hline\hline
\end{tabular}
\caption{ATLAS searches for light charged scalar (taken from Ref.~\cite{Coutinho:2024zyp}).}
\label{tab:ATLAS_charged}
\end{table}

\cleardoublepage
\section{}\label{app:C}
\begin{figure}[!h]
     \centering
             \includegraphics[clip=true,width=0.8\columnwidth]{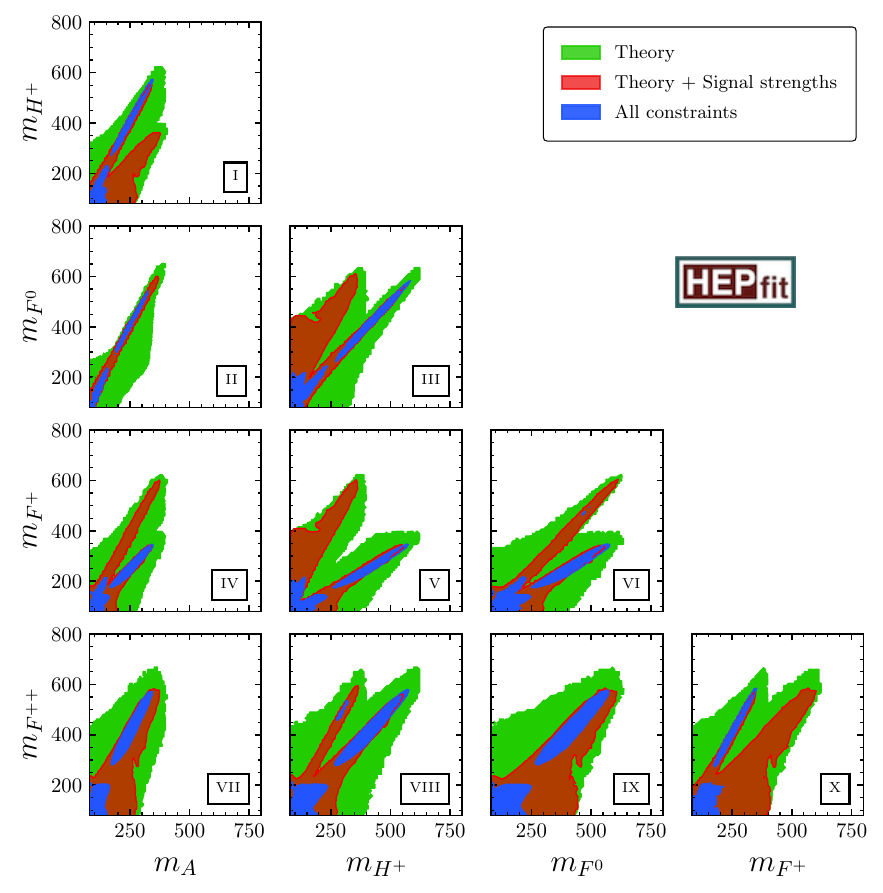}
     \caption{Allowed regions in the mass planes in the eGM model. The green, red, and  blue regions have the same meaning as in Figure~\ref{fig:6}.}
     \label{fig:9}
\end{figure}

In this appendix, we provide the supplementary figures for the eGM model. The allowed regions in the mass planes are shown in Figure~\ref{fig:9}. The colours in Figure~\ref{fig:9} have the same meaning as in Figure~\ref{fig:6}. In panels II and VII of Figure~\ref{fig:9}, we see that the combined fit reveals a mass hierarchy $m_{F^0}>m_A$ and $m_{F^{++}}>m_A$ in the relatively high mass region. In contrast, there is no preferred mass hierarchy among the remaining scalar states.  
\pagebreak
\bibliographystyle{JHEP}
\bibliography{LightHiggs}
\end{document}